\documentclass[10pt]{article}
\pdfoutput=1

\usepackage{jheppub} 
\usepackage{slashed}
\usepackage[T1]{fontenc}
\usepackage[latin9]{inputenc}
\usepackage{amsmath}
\usepackage{amssymb}
\usepackage{esint}
\usepackage{lmodern}
\usepackage{slashed}
\usepackage{dsdshorthand}
\usepackage{tikz} 
\usepackage[bottom]{footmisc}
\usepackage{mathdots}
\usetikzlibrary{arrows.meta} 
\usepackage{tabularx}
\usetikzlibrary{arrows}
\def\a{\alpha}

\def\vf{\varphi}

\def\bz{\bar{z}}

\def\thetab{\bar{\theta}}

\def\phid{\phi^{\dagger}}
\def\psid{\psi^{\dagger}}
\def\zb{\bar{z}}
\def\thetab{\bar{\theta}}

\def\veps{\varepsilon}
\def\Phid{\Phi^{\dagger}}
\def\vf{\varphi}
\def\vfd{\varphi^{\dagger}}

\def\td{\text{d}}
\def\fsi{\psi}

\newtheorem*{theorem*}{Theorem}



\title{Supersymmetry in the Landau Level Problem with a Disorder Potential}

\author[a,b]{Apratim Kaviraj}
\author[a,c]{and Philine van Vliet}

\affiliation[a]{Deutsches  Elektronen-Synchrotron  DESY,  Notkestr.   85,  22607  Hamburg,  Germany}
\affiliation[b]{Department of Physics, Indian Institute of Technology - Kanpur, Kanpur 208016,  India}
\affiliation[c]{Laboratoire de Physique Th\'eorique de l'\'Ecole Normale Sup\'erieure, PSL University,\\CNRS, Sorbonne Universit\'es, UPMC Univ. Paris 06, 24 rue Lhomond, 75231 Paris Cedex 05, France}

\emailAdd{philine.vanvliet@phys.ens.fr}
\emailAdd{akaviraj@iitk.ac.in}

\abstract{We explore a supersymmetric (SUSY) theory that arises in the Landau level problem with disorder. Charged particles in a strong magnetic field and a local potential are described by small excitations around the ground state, the lowest Landau level (LLL). Around forty years ago Br\'ezin, Gross and Itzykson showed that for a disorder potential certain observables in the LLL limit are described by a chiral SUSY theory. As a consequence, the problem undergoes a dimensional reduction by two dimensions, simplifying the computations. We generalize their findings by identifying the chiral SUSY as a special limit of a `magnetic Parisi-Sourlas (PS)' theory. The latter is a modification of the theories associated to fixed points of random field models. We show that the SUSY features extend to a much larger class of observables and to higher Landau levels. Finally we identify a set of new super-Ward identities and show how observables in the disordered theory must satisfy them.}

\begin{document} 

\maketitle

\section{Introduction} \label{sec:intro}

Disordered quantum field theories have many interesting physical applications. They describe systems with impurities, and are thus closely connected to setups realized in experiments, where samples are often not perfectly homogeneous. 
The impurities can be modeled by an interaction of pure degrees of freedom with a disorder field whose value at each coordinate is drawn from a distribution. If the impurities are very slowly fluctuating i.e. not in thermal equilibrium with the pure degrees of freedom, then the physical observables are computed for a given disorder field and then averaged over using the distribution. The observables obtained this way are called \emph{quenched averaged observables} and they correctly describe the large distance properties of an impure system. { The opposite case, where the impurities are in thermal equilibrium with the pure degrees of freedom, is called \emph{annealed disorder}.
Quenched disorder systems are in general more difficult to solve than annealed, and one has to use special techniques like replica trick to deal with them.} 

\paragraph{Random field models and Parisi-Sourlas supersymmetry:} Models with a quenched disorder have a lot of applications in statistical physics. An important area of interest is disordered critical points, i.e. when impurities change the critical behavior of a system and introduce new critical points described by disorder averaged observables.
Spin glasses are described by quenched disordered systems \cite{Edwards1975,spin-glass-book}, giving rise to a new type of ordered state: the `glassy' phase. Quenched disorder is also studied in connection to physics of polymers (see \cite{PhysRevA.20.2130,Vanderzande_1998}, and chapter 9 of \cite{Cardy-book}) and interfaces \cite{Wiese-notes}.

\begin{figure}[h]
	\centering \includegraphics[width=300pt]{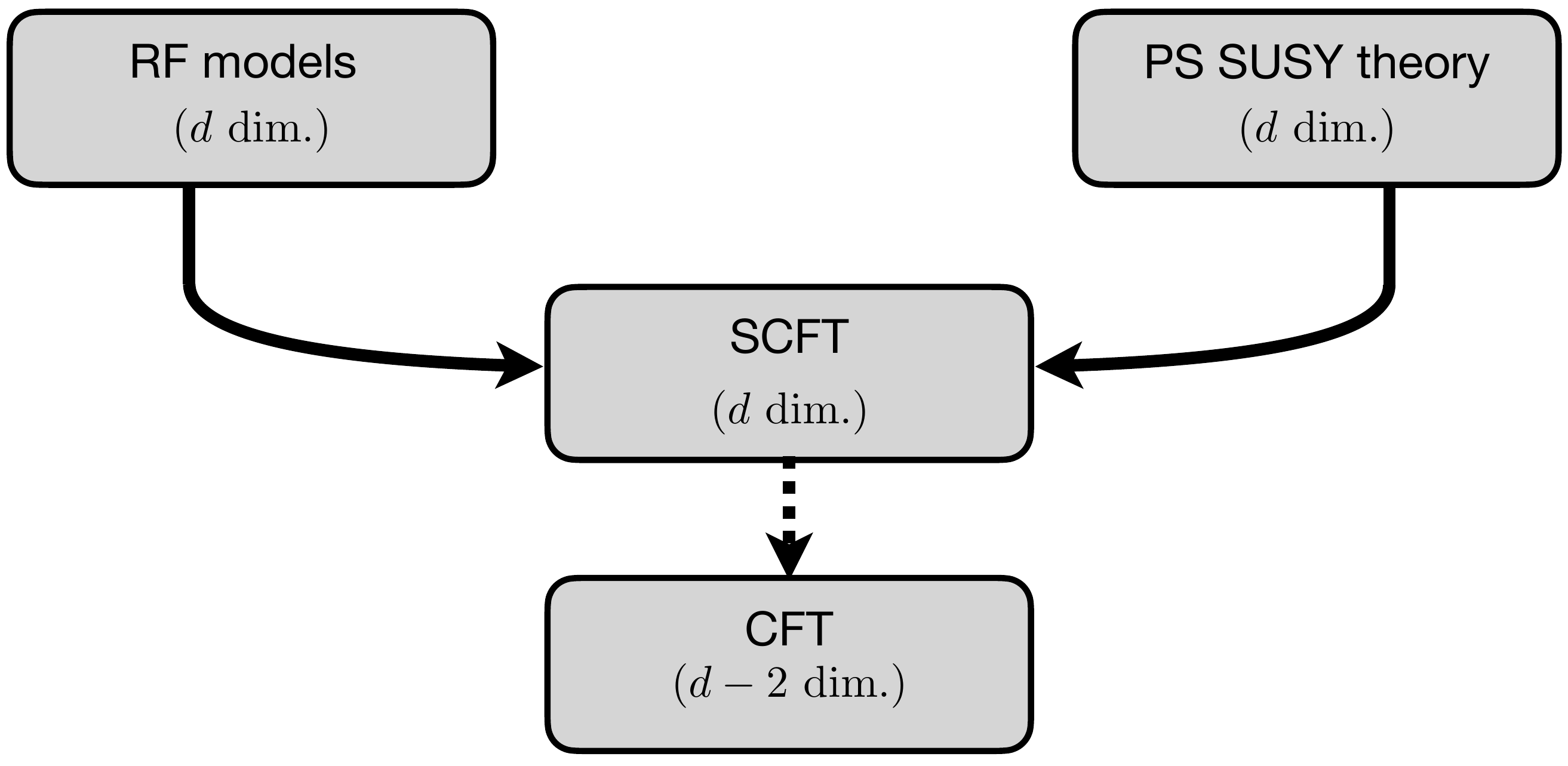}
	\
	\caption{\label{FigRFmodel} A summary of the Parisi-Sourlas conjecture on disordered RF model critical points. The curved arrow on the left indicates a non-SUSY RG flow in the RF model. The curved arrow on right denotes a SUSY-preserving RG flow of a Parisi-Sourlas theory. Both theories flow to the same SCFT$_d$ in the IR. As a consequence of PS SUSY an RF critical point has a dimensional reduction to a non-SUSY CFT$_{d-2}$.}
\end{figure}

A particularly interesting case of a disordered critical point is that of the special family of disordered models called random field (RF) models, see \cite{Rychkov:2023rgq} for a recent review. The most commonly studied examples are the RF Ising \cite{Picco1,Picco2,Picco3,Kaviraj:2020pwv, Kaviraj:2021qii, Wiese:2021qpk, 2020PhRvE.102f2154B, Tarjus_2016,Hikami:2018mrf, Piazza:2024wll} and the RF $\phi^3$ models \cite{PhysRevLett.46.871,zbMATH02068689,Hikami:2017sbg,Kaviraj:2022bvd}. It can be shown that these models have a critical point where the critical exponents are identical to those of the pure version of the same model in two dimensions less. This mysterious dimensional reduction property  was explained by Parisi and Sourlas \cite{Parisi:1979ka} with a conjecture that the critical point is described by a supersymmetric (SUSY) conformal field theory (CFT). 
The conjecture can be understood as follows (see also Fig. \ref{FigRFmodel}): 
\begin{itemize}
    \item The Parisi-Sourlas SUSY CFT (SCFT) emerges as an IR critical point in the RG flow of RF models. Along the RG flow SUSY is absent. 
    
    \item The Parisi-Sourlas SCFT is identical to the fixed point of a Parisi-Sourlas SUSY QFT (SQFT) in $d$-dimension. The SQFT has an OSp$(d|2)$ symmetry. It is is a simple extension of the usual Poincare symmetry, which is enhanced to a superconformal symmetry at the fixed point. The SUSY RG flow is different from that of the RF model.
    
    \item As a consequence of SUSY a large sector of correlation functions exhibits \emph{dimensional reduction}: when restricted to a $\hat{d}:=d-2$ dimensional submanifold, they become identical to those of a $\hat{d}$-dimensional local CFT, which describes the critical point of the pure model in $\hat{d}$-dimensions.
    
\end{itemize}
Let us briefly introduce the class of SUSY theories that appears in the above conjecture, which we will refer to as `Parisi-Sourlas (PS) SUSY theories' or simply `PS theories'. They can be  described with a scalar superfield $\Phi(x_i,\theta_1,\theta_2)$ defined in a superspace parameterized by $(x^i,\theta_1,\theta_2)$, with the generic form
\be\label{PS-susy}
S_{\text{PS}}=\int \td^dx\td \theta_2 \td\theta_2\, \Big[ \frac{1}{2}|\partial_a\Phi|^2+V(|\Phi|) \Big]\,.
\ee
Here $a$ is a superspace index and $V$ a generic interaction  - the other details of the action will be given elaborately in the main text. 
For the RF Ising and $\phi^3$ critical points $\Phi$ is real. Eq. \eqref{PS-susy} covers the slightly more general case of complex $\Phi$, which is more relevant for this paper. 

A PS theory has super-Poincar\'{e} invariance, i.e. symmetry under translations and rotations in superspace. A consequence of this symmetry is the dimensional reduction property
\be\label{dimred-stmnt}
\langle F(\Phi(\hat{x}_i))\rangle_{\text{PS}}=\langle F(\phi(\hat{x}_i))\rangle_{\widehat{d}}
\ee
where the l.h.s. is a correlation function of $\Phi(x_i)$'s in the PS theory $S_{\text{PS}}$ restricted to a $d-2$ hyperplane parameterized by $\hat{x}_i$, while the r.h.s. is a correlation function of a $\widehat{d}$-dimensional theory with its action identical to that of $S_{\text{PS}}$ with the replacement $\Phi\to \phi$, where $\phi$ is a non-supersymmetric scalar field. The statement can be proved to an arbitrary order in perturbation theory - see App. A of \cite{Kaviraj:2019tbg}, and \cite{PSproof2,CARDY1985123} for more rigorous proofs. 

It is possible to tune the parameters in $V(|\Phi|)$ to reach a critical point in the PS theory, where at each point along the RG flow SUSY is present. 
At the critical point the symmetry is extended to include superconformal invariance. For a Parisi-Sourlas SCFT, dimensional reduction can be shown to hold axiomatically at the level of operator product expansions and conformal blocks, without explicitly requiring a Lagrangian \cite{Kaviraj:2019tbg}. Since RF models flow to the same fixed point they possess SUSY and dimensional reduction at criticality, although these features are absent along the disordered RG flow. The presence of SUSY and dimensional reduction in RF critical points makes computations delightfully interesting and simple. \\

\paragraph{SUSY in disordered electron gas:} In this paper we will explore another physical example where a Parisi-Sourlas QFT emerges, this time in a nonrelativistic, non-conformal system. We consider the classic Landau quantization problem of charged particles, e.g. electrons, moving in a magnetic field. We will loosely refer to this system as `electron gas'. The charged particles do not interact with each other but see a local potential $h(x)$. The dynamics of a single such particle in $2d$ is given by the Hamiltonian
\be\label{eq:ham}
H=\frac{1}{2m}(p_i-e\,A_i)^2-\frac{e\,B}{2m} +h(x)\,.
\ee
Here $e$ is the charge of each particle of mass $m$. The coordinates are indexed with $i=1,2$. The external gauge field $A_i$ is a magnetic vector potential. The magnetic field is perpendicular to the 2d plane and its amplitude is given by $B=\varepsilon_{ij}\partial_i A_j$\,. When $h(x)=0$ the energy spectrum can be easily determined and is given by harmonic oscillator or Landau levels. We added a constant $-\frac{eB}{2m}$ to the Hamiltonian such that the ground state has zero energy. Then the $n$-th Landau level energy is
\be\label{Landaulevels}
E_n=\frac{n\, e\, B}{m}\,.
\ee
Each Landau level has an infinite degeneracy that is lifted when we turn on the interaction $h(x)$. This interaction could for example be an external electric potential. We will not specify the exact interaction, but we will demand that it is disordered.

We may write the partition function of the model as  $\text{Tr}\, e^{-\beta H}$, where $\beta$ is inverse temperature. Then the ratio $\beta B/m$ controls the gap between energies. If we set $B\to \infty$, or alternatively $m\to 0$, the first excited state $(n=1)$ gets widely separated from the ground state $(n=0)$. The electron is then restricted to the ground state, or the lowest Landau level (LLL), since it is much more energetically favorable. The restriction that the particle is in the ground state fixes its wave-function up to a holomorphic function $u(z)$, with $z=x_1+ix_2$. 

Let us briefly mention some of the major areas where the physics of a charged particle in a magnetic field becomes important. First and foremost, it is the foundation of the quantum Hall effect (QHE) and related phenomena (see \cite{tong2016lecturesquantumhalleffect} for a review). The presence of disorder plays a key role in localization of quantum Hall states \cite{PhysRevLett.42.673}. In the integer QHE (IQHE) this explains the characteristic plateaus (characterized by an integer $\nu$) in the dependence of conductivity on $B$. The fractional QHE (FQHE) is a harder problem, since there are new plateaus corresponding to fractional values of $\nu$ which can only be explained by including electron-electron interactions. See e.g. \cite{Jain_2007, Fradkin_2013} and references therein for more details on these broad topics. There is also a great deal of theoretical interest in the ground state of the Landau level problem, because LLL wave functions of magnetized particles have interesting connections with  correlation functions of a $2d$ free boson CFT \cite{MOORE1991362} or of $1/2$-BPS operators in $\mathcal{N}=4$ Super Yang-Mills theory on $S^3$ \cite{Corley:2001zk, Berenstein:2004kk}. 
\\

For the single electron problem an important observable to compute in the presence of disorder is the averaged Green's function $G(r,r'|E):=\overline{\langle r|(E-H)^{-1}|r'\rangle}$  where the bar denotes disorder average (this quantity is  related to the density of states per unit area - see section 20.4 of \cite{Wegner:2016ahw}). Now, a nice way to compute it is to map it to a 2-point correlation function of a field theory. This was first investigated by Wegner in \cite{Wegner83}, where it was shown from topological properties of Feynman diagrams in the disordered theory that the 2-dimensional problem becomes a zero dimensional one when it is restricted to LLL. 

These simplifications were explained by Br\'{e}zin, Gross and Itzykson \cite{BREZIN198424} as arising from an emergent supersymmetry. They showed that in the symmetric gauge the LLL limit of the theory can be repackaged in terms of a pair of chiral and anti-chiral superfields. The SUSY appearing here is similar to Parisi-Sourlas theories \eqref{PS-susy} because in perturbation theory the 2-point function computation also has a dimensional reduction from a 2d $\to$ 0d theory.  However this beautiful connection works only for a single observable: the LLL Green's function, and it is hard to extend it more general ones. The observation  of \cite{BREZIN198424} was proven nonperturbatively in \cite{KLEIN1985199}, and also used in \cite{KUNZ1986347} to derive further results. 

\begin{figure}[h]
	\centering \includegraphics[width=300pt]{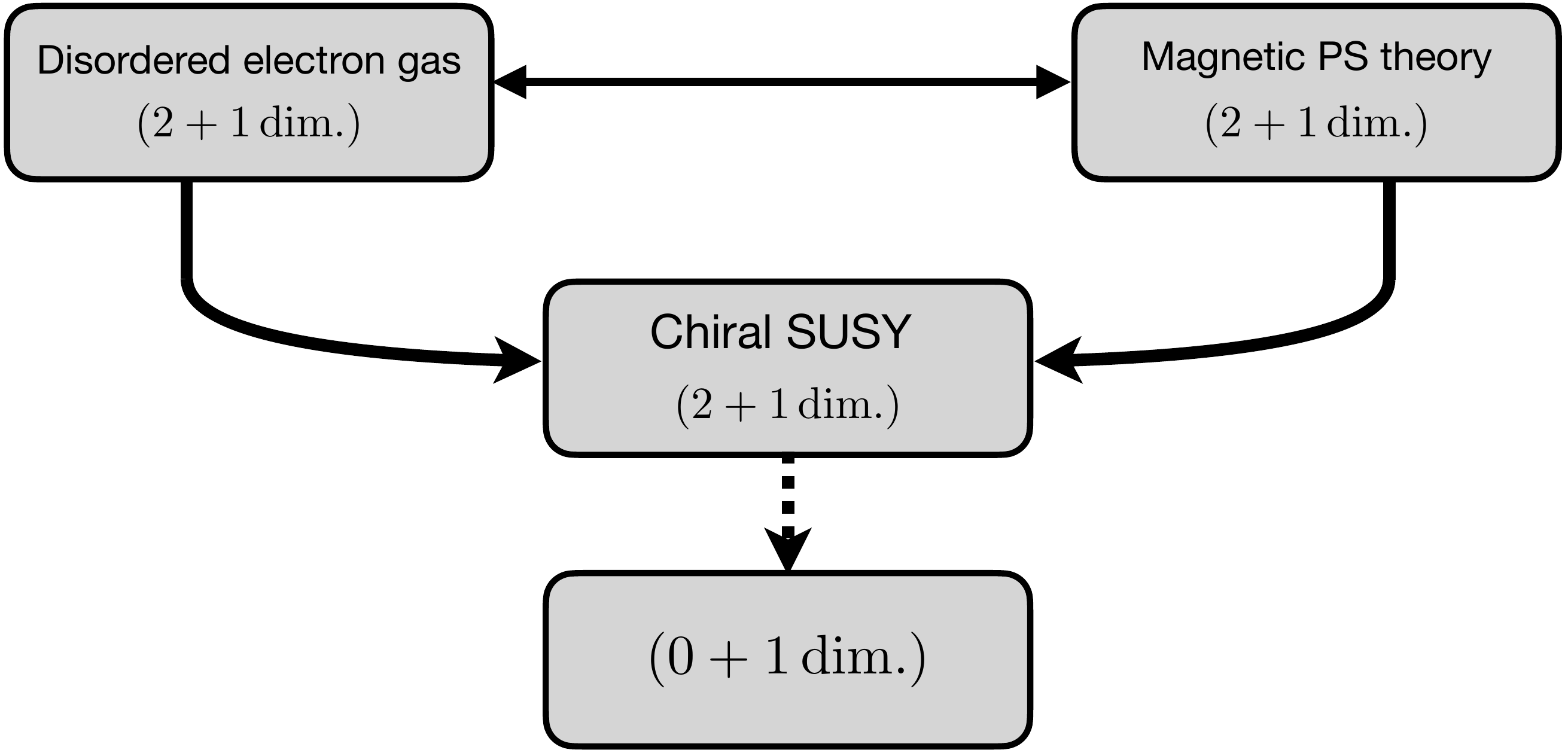}
	\
	\caption{\label{Figresults} A flowchart summarizing the SUSY connections of the disordered electron gas model. The left curved arrow indicates the LLL limit $m\to 0$ which was shown in \cite{BREZIN198424}. The right curved arrow indicates a similar limit in the (proposed) magnetic PS theory. The dotted arrow stands for dimensional reduction of the chiral SUSY theory (at $m\to 0$) to a $0+1$d QFT. The top left-right arrow denotes a map that connects the magnetic PS theory to the disorder problem even for nonzero $m$.}
\end{figure}

\paragraph{Goal and summary of results:}
In the analysis of \cite{BREZIN198424} the emergence of SUSY is sensitive to two things: 
\begin{enumerate}
     \item the LLL limit. \label{cond1}
    \item the choice of symmetric gauge. \label{cond2}
\end{enumerate}
Under these two conditions, and with the help of the chiral SUSY, the authors of \cite{BREZIN198424} were able to compute $G(r,r'|E)$.
However, there are certain observables whose definitions require one or both of these conditions to be relaxed, and they cannot be computed with the help of PS SUSY. As we will see further on in the paper, correlators of stress tensors or gauge currents are examples of such observables.
A natural question to ask is if the emergent SUSY theory can be realized as a limit of a more general Parisi-Sourlas QFT. 
We will show in this paper that the latter exists and that it can be mapped to the disordered electron gas even when condition \ref{cond1} or \ref{cond2} is relaxed. Such a map unavoidably involves SUSY-breaking deformations.
Nevertheless, it can still be used to derive constraints coming from SUSY on a more general class of observables in the disordered theory. These findings are summarized in Fig. \ref{Figresults}.

In this paper, we present a `magnetic Parisi-Sourlas' theory that is a usual PS theory coupled to an external `super'-gauge field which is related to a `super'-magnetic field. We show how, with an appropriate tuning of the super-gauge field, we can relax both the conditions \ref{cond1} and \ref{cond2}, or one of them individually. Furthermore, 
using the magnetic PS framework we identify a larger class of observables for which dimensional reduction will hold. Finally, we are able to formulate a set of super-Ward identities, and we point out how these identities must hold for the disordered theory observables in the LLL limit. 

In section \ref{sec:model} we introduce the model of noninteracting particles in a magnetic field as studied by \cite{BREZIN198424}, and show the emergent Parisi-Sourlas supersymmetry by using replica fields. We recast the constraints for the particles to be in the lowest Landau level in a slightly different form than in \cite{BREZIN198424}, following \cite{Son:2013rqa}, which allows us to write down the currents and the stress-tensor, and the corresponding Ward identities. 
In section \ref{sec:gensusy}, we dive deeper into the emergent SUSY and show how to obtain the same field theory starting from a supersymmetric, non-relativistic QFT. We derive the LLL conditions for superfields, which emerge from the equations of motion for both bosonic and fermionic degrees of freedom. We also show how dimensional reduction is achieved when taking the LLL limit, for which the mass of the electron $m$ is taken to zero. 
Then, in section \ref{sec:Wardid}, we derive supercurrents, a super-stress-tensor, and corresponding super-Ward identities. 
We show how they reduce to the ones found in section \ref{sec:model} when taking the LLL limit.
We finish with a summary and outlook on future applications of our work in section \ref{sec:outlook}. There are four appendices for  additional details on some of the important points in the main text. 

\section{Particle in a magnetic field}\label{sec:model}
\subsection{The model}
Consider a system of charged particles in an external gauge field and a potential whose Hamiltonian is given by eq.~\eqref{eq:ham}. Their dynamics can be described by the following $2+1$ dimensional nonrelativistic field theory of a complex spinless (i.e. scalar) field:
\be\label{eq:org-dis}
S[\phi,h]=\int \text{d}^2x \text{d}t \Big(i \phid D_0  \phi -\frac{1}{2m}|D_i \phi|^2 +\frac{B}{2m}\phid\phi+ h(t,x)\,\phid\phi\Big)
\ee
Here, $D_0=\partial_0-i A_0$ and $D_i=\partial_i-i A_i$ respectively denote the covariant derivatives, and we have identified $x_0 \equiv t$. We will call~\eqref{eq:org-dis} the disordered electron gas model. 

For simplicity we consider $\phi$ to be bosonic which corresponds to a system of bosonic particles. For fermionic statistics (e.g. a system of electrons)  we should take $\phi$ to be a complex spinless fermion. It is easy to modify our analysis that follows for fermions, and in fact most of our findings will be indifferent to this, see App. \ref{app:fermion-dis}. Note that \cite{BREZIN198424} also started with a disordered bosonic action since their focus was a single electron Hamiltonian for which boson/fermion statistics does not matter. \\
The theory is invariant under the gauge transformations
\be
\phi\to e^{-i\alpha}\phi\,, \ A_\mu\to A_\mu+\partial_\mu \alpha\,.
\ee
We consider the disorder field $h(t,x)$ to be drawn from a distribution $\mathcal{P}(h(x))$ whose mean and variance are respectively given by :
\begin{align}
&\overline{h(t,x)}:= \int \mathcal{D}h\, \mathcal{P}(h)\, h(t,x)=0\,,\\
&\overline{h(t,x)h(t',x')}=2\,\lambda\, \delta(x-x')c(t-t')\,.\label{eq:avgdef}
\end{align}
The higher moments i.e. $\overline{h(t_1,x_1)\cdots h(t_n,x_n)}$\, with $n>2$ are given in terms of the variance by Wick's theorem. The constant $\lambda$ is the strength of the disorder. The factor $c(t-t')$ should be set to 1 for time-independent white noise disorder. But we will see in the next subsection that choosing it to be a delta function gives a local form to the interaction when we average over the disorder. 

We are interested in the averaged observables $\overline{\langle A(\phi)\rangle}$ in this theory, where $\langle A(\phi)\rangle$ denotes some correlation function of fields $\phi$ or local composite operators built out of $\phi$'s. E.g. $A(\phi)=\phi(x)\phi(0)$ gives a two point function.

We consider the disorder to be \emph{quenched}. This means the fluctuations of the impurities are not considered dynamic degrees of freedom. Observables are measured for a fixed configuration of the disorder and then averaged over: 
\footnote{One can think of this by taking a very large impure system and then dividing it into smaller (yet large) regions/intervals. The potential in each subsystem is fixed but is uncorrelated to other subsystems. Then an averaged observable could correspond to measuring that observable in various subsystems and performing an average. We may only focus on such self-averaged observables i.e. whose values are given by \eqref{quenchavg}.}
\be\label{quenchavg}
\overline{\langle A(\phi)\rangle}=\int \mathcal{D}h\, \mathcal{P}(h)\, \langle A(\phi) \rangle\,.
\ee
For quenched averaged observables a nice way to proceed is to use replica trick that we introduce in the next subsection \ref{repac}. The reader familiar with replicas may skip to  subsection \ref{rep-ferm} where we map replica fields to fermions by Gaussian integral identities. 

\subsection{Replica theory}\label{repac}

Instead of working directly with the action~\eqref{eq:org-dis}, it would be preferable to have an action that does not depend on $h(t,x)$ and effectively gives us the quenched averaged observables directly. Since $h$ is not a dynamic field we cannot simply average over it in the partition function by the path integral. However, there is a way to find such an effective action using the so-called \emph{replica theory}. 
Let us start with the quenched averaged observable, and write it in the following way: 
\begin{align}
\overline{\langle A(\phi)\rangle}&=\int\mathcal{D}h \, \mathcal{P}(h)\,  \frac{\int \mathcal{D}\phi \, A(\phi)\, e^{-S[\phi,h]}}{\int \mathcal{D}\phi \, e^{-S[\phi,h]}} \label{correlator}\\
&= \int\mathcal{D}h \,  \mathcal{P}(h) \,  \frac{\int \mathcal{D}\phi_1\cdots\mathcal{D}\phi_n \, A(\phi_1)\, e^{-\sum_{\a=1}^n S[\phi_\a,h]}}{\big(\int \mathcal{D}\phi \, e^{ -S[\phi,h]}\big)^n}
\end{align}
In the numerator and denominator inside the $h$ integral, we have inserted  a path integral of $n-1$ fields (replicas) with the action $\sum_{\a=1}^n S[\phi_{\a},h]$. In the numerator we call these fields $\phi_\a$ with $\a=2,\cdots, n-1$ being the `replica index', and we defined $\phi_1:= \phi$. The denominator is a partition function of $n$ decoupled replica theories.

While computing $\langle A(\phi_1)\rangle$ in the replica theory, we may treat $n$ as a continuous parameter. In the limit $n\to 0$ the denominator becomes 1. Now we are allowed to integrate $h(t,x)$ along with the $\phi_\a$ path integral:
\begin{align}
\overline{\langle A(\phi) \rangle}=& \lim_{n\to 0}\int \prod_{\a=1}^n \mathcal{D}\phi_\a\mathcal{D}h \,  \mathcal{P}[h]\, A(\phi_1)\, e^{-\sum_{\a=1}^n S[\phi_{\a},h]}\nonumber\\
=&\lim_{n\to 0}\int \prod_{\a=1}^n\mathcal{D}\phi_\a\, A(\phi_1)\, e^{-S_{\text{rep}}[\phi_\a]}\label{eq:dis-corr-repl}\\
S_{\text{rep}}[\phi_\a]=& \int \text{d}^2x \text{d}t\left[i \sum_\a \big({\phid_\a}D_0\phi_\a - \frac{1}{2m} |D_i\phi_\a|^2+\frac{B}{2m}\phid_\alpha\phi_\alpha\big)+ \lambda\Big(\sum_\a\phi_\a^\dagger \phi_\a\big)^2\right] \label{eq:rep-action}.
\end{align}
In the last step we have used the disorder properties \eqref{eq:avgdef}. For simplicity we have chosen $c(t-t')=\delta(t-t')$ in order to have a local interaction.\footnote{Other choices would result in a nonlocal quartic interaction, e.g. $c(t-t')=1$ leads to $\int \td t\, \td t'\phid\phi(t)\phid\phi(t')$ in \eqref{eq:rep-action} after  disorder averaging. This will not change the analysis of this and the next subsection by much (see footnote \ref{footnote-G}). It will however make the supersymmetric framework of sec. \ref{sec:proj-LLL} more nontrivial.}
The replica action~\eqref{eq:rep-action} describes our model with a disorder potential. Notice that averaging out the disorder has led to a quartic term with coupling $\lambda$\,, that couples all the replica fields.

We can generalize \eqref{eq:dis-corr-repl} to include averages of products of any number of correlators. If we have for example two correlators $\langle A(\phi)\rangle$ and $\langle B(\phi)\rangle$, their disorder average is given by
\be\label{avg-repl-2}
\overline{\langle A(\phi)\rangle\langle B(\phi)\rangle}=\lim_{n\to 0}\int \prod_{a=1}^n\mathcal{D}\phi_a\, A(\phi_1)B(\phi_2)\, e^{-S_{\text{rep}}[\phi_\a]}\,.
\ee

\subsection{From replicas to global OSp}\label{rep-ferm}
We have shown how a computation using the replica action~\eqref{eq:rep-action} gives us a quenched averaged correlation function in the limit of zero replicas. We now show that the $n\to 0$ limit of the theory can be mapped to a theory of scalar bosons and fermions. It will turn out that this new theory has an underlying local supersymmetry in the massless limit $m\to 0$. 

The quartic term in~\eqref{eq:rep-action} can be written as a Gaussian term by introducing an auxiliary field $\xi(x,t)$ in the action:\footnote{\label{footnote-G}We may consider a more generic interaction $G\big(\sum_\a \phid_\a \phi_\a(x,t)\big)$ where $G(\alpha)$ can be a polynomial or even a nonlocal interaction. E.g. we will have $G(\alpha)=-\int \td^2x\td t\td t'\alpha(x,t)\alpha(x,t')$ for a time-independent white noise disorder\,. If we have the exponential $\exp\big[G({\sum} \phi^{\dagger}_{\a} \phi_{\a})\big]$
we may obtain it from a quadratic Lagrangian of replicas by introducing two auxiliary fields $\xi_1,\xi_2$ and an action $\exp \big[(G(\xi_1)+\int \xi_2(\xi_1-\sum \phi^{\dagger}_{\a} \phi_{\a}))\big]$ where $\xi_2$ is a Lagrange multiplier that equates $\xi_1=\phid_\a\phi_\a$. This allows the introduction of fermions as in \eqref{PIident}. Also in \eqref{eq:bos-fer} one should appropriately replace the local quartic interaction  with $G(\varphi^\dagger\varphi+\psid\psi)$.}
\be\label{eq:replica-aux}
S_{\text{rep}}[\phi_\a,\xi]= \int \text{d}^2x \text{d}t\left[ \sum_\a \big(i{\phid_\a}D_0\phi_\a  - \frac{1}{2m} |D_i\phi_\a|^2+\frac{B}{2m}\phid_\a\phi_\a+i \xi\phid_\a\phi_\a\big)+\frac{\xi^2}{4\lambda} \right]\,.
\ee
We now focus on the path integral of the replicas $\phi_a$\,, $a=2,\cdots , n$. 
It is a Gaussian path integral over $n-1$  complex variables that allows us to make the identification
\be\label{PIident}
\int\, \prod_{\a=2}^n\mathcal{D}\phi_\a\exp\big[- \sum_{\a=2}^{n}\phid_\a \,\text{M}\, \phi_\a\big] \ =\left(\frac{\text{det M}}{\pi}\right)^{1-n} \ \  \stackrel{n\to 0}{=}\int\mathcal{D}\psi_1\mathcal{D}\psi_2\, \exp\big[- i \psi_1 \text{M} \psi_2\big]\,.
\ee
Here M$=i D_0 + \frac{1}{2m}D_i^2+ i \xi $. At $n\to 0$ the path integral is equivalent to that of anticommuting variables $\psi_1,\psi_2$, two real scalar fermions. For later convenience we pack them together by defining a complex fermion and its conjugate:
\begin{align}
\psi&= \frac{1}{\sqrt{2}} (\psi_1+i\,\psi_2)\,,\nonumber\\
\psid&= \frac{1}{\sqrt{2}} (\psi_1-i\,\psi_2)\,.
\end{align}
This maps $\psi_1 \text{M} \psi_2 \to - \frac{i}{2} \psid \text{M} \psi$\, +\, \text{c.c.} up to total derivative terms\,.
We may now re-insert the resulting quadratic fermion term back in \eqref{eq:replica-aux}.
By integrating out the auxiliary field $\xi$ and introducing $\varphi:=\phi_1$  we map $S_{\text{rep}}[\phi_\a,\xi]$, to the new action
\be\label{eq:bos-fer}
S[\varphi,\psi]= \int \text{d}^2x \text{d}t\left[i {\varphi^\dagger}D_0\varphi + i \psid D_0\psi - \frac{1}{2m}( |D_i\varphi|^2+|D_i\psi|^2)+ \frac{B}{2m}(\varphi^\dagger\varphi+\psid\psi)+\lambda(\varphi^\dagger\varphi+\psid\psi) ^2\right]\,.
\ee
This action has an underlying global OSp$(2|2)$ symmetry which is simply a consequence of the $n\to 0$ limit of the $U(1)^n\times O(n)$ symmetry of replicas \cite{refId0}. The global symmetry is realised under the general field transformations (denoting $\text{Re }\vf=\vf_1$ and $\text{Im }\vf=\vf_2$)
\begin{align}
&\vf_i\to a_{ij} \vf_j + b_{ij} \psi_{j}\\
&\psi_i\to c_{ij}\vf_{j} + d_{ij}\psi_{j}\,,
\end{align}
with the parameters $a,d$ ($b,c$) being Grassmann even (odd) and preserving the quantity $\vf^\dagger\vf+\psid\psi$\,.  Notice that a subgroup of the above is a global Sp$(2)$ that preserves the $\psi,\psid$ part of the action. We will refer to $S[\vf,\psi]$ as the `global OSp theory'. 

The path integral identity \eqref{PIident} leads to certain equalities in correlation functions. Indeed singlets of the global Sp$(2)$ built from the fermions can be mapped to the singlets of U$(1)^{n-1}\times$O$(n-1)$ subgroup of the replica theory. We discuss this in more detail in App. \ref{app:map} (see also appendix C of \cite{Kaviraj:2020pwv}). In particular it is straightforward to show that correlation functions of the composite operator $\psid\psi$ are equivalent to those of $\sum_{\a=2}^n\phid_\a\phi_\a$ as $n\to 0$. For example:
\be
\langle (\psid\psi)(x)\, (\psid\psi)(y)\rangle= \lim_{n\to 0}\langle \sum_{\a=2}^n\phid_\a\phi_\a(x)\sum_{\a=2}^n\phid_\a\phi_\a(y)\rangle\,.
\ee
This map can be generalized to bilocal operators, e.g. Re\,$\sum_{\a=2}^n\phid_\a(x)\phi_\a(y)$ for $n\to 0$ can be related to Re\,$\psid(x)\psi(y)$ in the respective path integrals. This allows us to map certain two- and higher-point correlation functions of the fermions to those of the replicas. In other words, we can use the equality
\be\label{eq:2point-triv}
\lim_{n\to 0}\sum_{\a=2}^{n}\,\text{Re}\,\langle \phi_\a(x)\phid_\a(0)\rangle = \text{Re}\,\langle \psi(x)\psid(0)\rangle\,.
\ee
There are similar relations for higher-point functions (see App. \ref{app:map}). \\ \\
Furthermore, using replica symmetry and~\eqref{eq:dis-corr-repl}, we can write 
\be
\lim_{n\to 0}\sum_{\a=2}^{n}\,\text{Re}\,\langle \phi_\a(x)\phid_\a(0)\rangle = \lim_{n\to 0}\sum_{\a=2}^{n}\,\text{Re}\,\frac{1}{n-1}\langle \phi_1(x)\phid_1(0)\rangle = - \text{Re}\,\overline{\langle \phid(x)\phi(0)\rangle}
\ee
This also sugggests the equality of the two point functions Re $\langle\psid\psi\rangle= - \text{Re}\, \langle\varphi^\dagger\varphi\rangle$ in \eqref{eq:bos-fer} since $\varphi=\phi_1$, reflecting the underlying global OSp$(2|2)$ symmetry. 

Let us point out that in \cite{BREZIN198424} the analog of the global OSp theory was obtained by rewriting the denominator of \eqref{correlator}, a Gaussian theory, as the path integral of fermions and then disorder averaging. A similar analysis for disordered electron gas was developed in \cite{Efetov:1983xg, Efetov:1997fw} where one obtains a nonlinear supersymmetric sigma model. Recently this was extended for SYK models \cite{Sedrakyan:2020oip}. 

\subsection{Projection to the lowest Landau level}\label{sec:proj-LLL}
Recall from sec. \ref{sec:intro} that we may take large $B$ or equivalently the limit $m\to 0$ in the action \eqref{eq:bos-fer} in order to restrict the problem close to its ground state.\footnote{As mentioned in sec. \ref{sec:intro} a few lines below~\eqref{Landaulevels}, the massless limit should be understood as setting $m/B\ll \beta$, while keeping $\lambda/\beta$ fixed. We may alternatively set $m/B\ll \lambda$ implying that the separation of higher Landau levels from the ground state is large compared to the (disorder) interaction.} We follow the steps of \cite{Geracie:2014nka} to take this limit smoothly without encountering a singularity. 
First let us introduce a vector $\vec\phi:=(\vf,\psi)$ to write the global OSp theory in \eqref{eq:bos-fer} more compactly:
\be\label{eq:ac-with-B}
S[\vec\phi]= \int \text{d}^2x \text{d}t\left[i ({\vec{\phi}^\dagger}\cdot {D_0}\vec\phi) -\frac{1}{2m}(D_i\vec\phi)^\dagger\cdot (D_i\vec\phi)+\frac{B}{2m}\vec\phid\cdot\vec\phi+\lambda(\vec \phi^{\dagger}\cdot \vec\phi) ^2\right]\,,
\ee
In above we denote $\vec \phid\cdot \vec \phi = \vfd\vf+\psid\psi$ and so on. Notice that from the definition of the covariant derivative $D_i$ one can write $B\, (f^\dagger f)=i\varepsilon^{ij}(D_i f)^\dagger (D_jf)$ up to total derivatives. 
This lets us write the action \eqref{eq:ac-with-B} as follows
\begin{equation}\label{scalar-EFT}
 S[\vec\phi] = \int \text{d}^2x \text{d}t\left[i ({\vec{\phi}^\dagger}\cdot {D_0}\vec\phi) - \frac{(\eta^{ij}+ i \varepsilon^{ij})}{2m} (D_i\vec\phi)^\dagger\cdot (D_j\vec\phi)+\lambda(\vec \phi^{\dagger}\cdot \vec\phi) ^2\right] .
\end{equation}
Here $\eta^{ij}=\delta^{ij}$ is the flat metric in 2d and $\varepsilon^{ij}$ an antisymmetric tensor with $\epsilon^{12} = 1$. Let us now consider a more general version by replacing $\eta^{ij} \to g^{ij}$, where $g^{ij}$ is an arbitrary (symmetric) metric in 2d. It will turn out to be useful to write this metric as a product of \emph{Zweibeine} $e^{a}_{i}$ in the following way:
\be
g_{ij} = \delta_{ab} e_{i}^a e_{j}^b\:, \quad e^{a}_i e^{bi} = \delta^{ab}\:.
\ee
The expression for $g_{ij}$ reduces to the flat metric $\eta_{ij}$ if we identify the Zweibeine with Kronecker deltas: $e_{i}^{a} = \delta_{i}^{a}$. We adopt this identification throughout the paper, since ultimately we are only interested in the flat limit. However, one could choose to keep the Zweibeine arbitrary. Later on it will be convenient to introduce complex Zweibeine: 
\be
e_i = \frac{1}{2} (\delta_{i}^{1} - i \delta_{i}^{2})\:, \quad \bar{e}_i = \frac{1}{2}(\delta_{i}^{1} + i \delta_{i}^{2})\:, \quad e^i e_i = \bar{e}^i \bar{e}_i = 0\:, \quad e^i \bar{e}_i = \bar{e}^i e_i = \frac{1}{2}\:,\label{eq:zweibeinedef1}
\ee
where we have already replaced the Zweibeine with Kronecker deltas. The metric can now be rewritten in the following way:
\be
g^{ij} = 2 \, (e^i \bar{e}^j + \bar{e}^i e^j)\:.
\ee
Similarly, we can express the antisymmetric tensor in terms of the complex Zweibeine as:
\be
\varepsilon^{ij} = - 2 i \, (e^i \bar{e}^j - \bar{e}^i e^j)\:.
\ee
From the above equations the identity $\eta^{ij}+ i \varepsilon^{ij}= 4 e^{i}\,\bar{e}^{j}$ follows.
We then write the covariant derivative terms as:
\be\label{eq:step-intro-X}
(\eta^{ij}+ i \varepsilon^{ij})\Big((D_i \vf)^{\dagger} D_j \vf+(D_i \psi)^{\dagger} D_j \psi\Big) \ = 4 \left( \ (\bar{e}^iD_i \vf)^{\dagger} (\bar{e}^jD_j \vf)+(\bar{e}^iD_i\psi )^{\dagger} (\bar{e}^jD_j \psi)\right)\,.
\ee
It is possible to rewrite these square terms using a Hubbard-Stratonovich transformation, which introduces the path integrals over two auxiliary field $\chi_1$ and $\chi_2$:
\begin{align}\label{eq:aux-fields}
&\exp\Big[-\int_{x,t}\, \frac{2}{m} \big( (\bar{e}^iD_i \vf)^{\dagger} (\bar{e}^jD_j \vf)+(\bar{e}^iD_i\psi )^{\dagger} (\bar{e}^jD_j \psi)\big)\Big]\nonumber\\
&=\int \mathcal{D} \chi_1\mathcal{D} \chi_2\, \exp\Big[- \int_{x,t} \big( \chi_1^{\dagger}(\bar{e}^iD_i \vf)+ \chi_1(\bar{e}^iD_i \vf)^{\dagger}+ \chi_2^{\dagger}(\bar{e}^iD_i \psi)+ \chi_2(\bar{e}^iD_i \psi)^{\dagger}- \frac{m}{2}\, (|\chi_1|^2+|\chi_2|^2) \big)\Big] \,.
\end{align}
Here $\chi_1$ and $\chi_2$ are complex bosonic and fermionic field respectively, so that all terms in the exponential are Grassmann even.  
Finally we introduce another 2-component vector $\vec \chi:=(\chi_1,\chi_2)$.  The global OSp theory may now be written as:
\be\label{eq:lagrangemult}
S[\vec \phi,\vec \chi]= \int \text{d}^2x \text{d}t\left[i\,  {\vec{\phi}^\dagger}\cdot {D_0}\vec\phi -\vec \chi^{\dagger}\cdot (\bar{e}^iD_i\vec\phi)+\text{c.c.}+\frac{m}{2}\, \vec \chi^{\dagger}\cdot \chi+\lambda(\vec \phi^{\dagger}\cdot \vec\phi) ^2\right]\,.
\ee
This action lets us take the $m\to 0$ limit smoothly. Varying the action with respect to $\vec \chi$ in the $m\to 0$ limit gives the equation of motion $e^i D_i \vec \phi=0$ , or in other words
\be\label{LLLcond}
\bar{e}^iD_i\vf=0\,, \quad \bar{e}^i D_i\psi=0\,.
\ee
These equations force the fields $\varphi, \psi$ to be holomorphic, and hence impose the lowest Landau level (LLL) conditions on $\vf$ and $\psi$.  Recall that the covariant derivative $D_i$ depends on $A_i$ which is a non-dynamic gauge field. We may choose the Lorentz gauge to fix $A_i= \frac{B}{2} \varepsilon_{ij}x^j$\,. 
If we now introduce the holomorphic and anti-holomorphic coordinates $z=x+iy\,, \zb=x-iy$ respectively, we get the LLL conditions in the familiar form :
\begin{align}\label{eq:lll-sp-def}
&D_{\bar{z}} \, \vf =D_{\bar{z}} \, \psi=0\,,\\ &D_{\bar{z}}:= \partial_{\bar{z}} + \frac{B}{4} z\,.
\end{align}
These have solutions in terms of holomorphic fields:
\be\label{lll-sol}
\vf(z,\bar{z},t)=e^{-\frac{B}{4}|z|^2}u(z,t)\,, \quad \psi(z,\bar{z},t)=e^{-\frac{B}{4}|z|^2}v(z,t)\,.
\ee
Here $u(z,t)$ and $v(z,t)$ are respectively bosonic and fermionic, and independent of $\zb$. The complex conjugate fields $\vfd,\psid$ can be solved in terms of anti-holomorphic fields $u^{\dagger},v^\dagger$ respectively.

The $m\to 0$ limit of~\eqref{eq:lagrangemult} is equivalent to enforcing the conditions~\eqref{LLLcond} in the path integral, i.e. an insertion of $\delta(e^iD_i\vf) \ \delta(e^iD_i\psi)\times (\text{c.c.})$. This amounts to replacing the path integral over $\vf$ and $\psi$ to $u(z,t)$ and $v(z,t)$ respectively. Upon doing this, we get the LLL theory
\be\label{eq:LLL-th}
S_{\text{LLL}}[u,v]=\int \text{d}z \text{d}\bar{z}\text{d}t \, \big[e^{-\frac{B}{2}|z|^2} i (u^\dagger D_0u+v^\dagger D_0v) +\lambda\big(e^{-\frac{B}{2}|z|^2}(u^\dagger u+v^\dagger v)\big)^2\Big]\,.
\ee

\subsection{Chiral supersymmetry at the lowest Landau level}
In this subsection we will closely follow the argument of \cite{BREZIN198424} and show that~\eqref{eq:LLL-th} has a local {`chiral'} supersymmetry. To do this let us first introduce a pair of complex Grassmann coordinates and a respective Grassmann integral as follows:~\footnote{\label{pifootnote}Note the factor of $\pi$ difference in conventions with respect to \cite{BREZIN198424}.}
\begin{align}
\theta=\theta_1 - i\theta_2\,,& \, \, \thetab=\theta_1 + i\theta_2\, \label{complex-Grassmann}\\
\int \text{d}\theta_1\text{d}\theta_2 \, \theta_1\theta_2=\frac{1}{4} & \implies\int \text{d}\theta \, \text{d}\thetab \, \thetab \theta=1\,.\label{complex-Grassmann-int}
\end{align}
Here $\theta_1=\theta_1^{*}, \, \theta_2=\theta_2^{*}$ are real Grassmann coordinates. The complex Grassmann coordinates $\theta,\thetab$ satisfy $\theta^{*} = \thetab, \, \thetab^{*}=\theta$ and the reality condition $(\theta\thetab)^{*}=\thetab^{*}\theta^{*}=\theta\thetab$. \\ \\ 
We also introduce a pair of chiral and anti-chiral superfields that respectively depend on $(z,\theta)$ and $(\zb,\thetab)$ only:
\begin{align}\label{hol-supfield}
U(z,\theta)&:=u(z)+\frac{k}{\sqrt{2}}\,\theta \, v(z)\, \ \implies\   e^{-\frac{k^2\theta\bar{\theta}}{4}}U(z,\theta)=u(z)+\frac{k}{\sqrt{2}}\,\theta\, v(z)-\frac{k^2\theta\thetab}{4}u(z)\,,\nonumber\\
U^{\dagger}(\zb,\thetab)&:=u^{\dagger}(\zb)+\frac{k}{\sqrt{2}}\,{v}^\dagger(\zb)\, \thetab\, \ \implies \   e^{-\frac{k^2\theta\bar{\theta}}{4}}U^{\dagger}(\zb,\thetab)=u^{\dagger}(\zb)+\frac{k}{\sqrt{2}}\,{v}^\dagger(\zb)\, \thetab-\frac{k^2\theta\thetab}{4}u^{\dagger}(\zb) \,.
\end{align}
Here $k:=\sqrt{B}$\,. The existence of the holmorphic superfield $U$ suggests that it could be obtained as a SUSY generalization of the LLL conditions on a more general local superfield. In the next subsection we will see how to make this statement more precise. 
\\
The superfields in~\eqref{hol-supfield} satisfy the identity
\be
u^{\dagger}f(D_0)u+v^{\dagger}f(D_0)v= \frac{2}{k^2} \int \text{d}\theta\text{d}\thetab \ e^{-\frac{k^2}{2}\theta\bar{\theta}}U^{\dagger}f(D_0)U
\ee
Using this identity, the action in~\eqref{eq:LLL-th} can be written as
\begin{align}\label{eq:LLL-Phi}
S[U]
= \int &\text{d}t\text{d}^2z \, \text{d}\theta  \text{d}\bar{\theta}  \,\big(e^{-\frac{k^2}{2}(|z|^2+\theta\thetab)} \, i U^\dagger D_0 U\big) 
+ \lambda\int \text{d}t\text{d}^2z  \,\bigg[e^{-\frac{k^2}{2}|z|^2} \frac{2}{k^2} \int \text{d}\theta \, \text{d}\bar{\theta} \, e^{-\frac{\kappa^2}{2} \theta\thetab}U^\dagger U\bigg]^2\,.
\end{align}
It was pointed out in \cite{BREZIN198424} that powers of $\theta,\thetab$ integrals, as in the last term of the second line above, have the following property 
\be
\bigg[\frac{2}{k^2}\int \text{d}\theta \text{d}\bar{\theta} \, e^{-\frac{k^2}{2} \theta\thetab}U^\dagger U\bigg]^n=\frac{2}{n\,k^2}\int \text{d}\theta \text{d}\bar{\theta} \, e^{-\frac{nk^2}{2} \theta\thetab}(U^\dagger U)^n\,,
\ee
and the second term of~\eqref{eq:LLL-Phi} can be simplified into a single $\theta,\thetab$ integral. We arrive at the final expression for the superfield action in~\eqref{eq:LLL-Phi}: 
\be\label{eq:susylll}
S[U]=\int \text{d}t\,\text{d}z \, \text{d}\bar{z}\, \text{d}\theta \, \text{d}\bar{\theta}  \, \, \Big[e^{-\frac{k^2}{2}(|z|^2+\theta\thetab)} i U^\dagger D_0 U + \lambda\, \big(e^{-\frac{k^2}{2}(|z|^2+\theta\thetab)} U^\dagger U\big)^2\Big]\,.
\ee
We will refer to $S[U]$ as a chiral SUSY theory since it is described by a pair of chiral and anti-chiral superfields. 

Let us remind the reader that the quartic superpotential in \eqref{eq:susylll} is specific to  the disorder \eqref{eq:avgdef} with $c(t)=\delta(t)$. For more general disorder, e.g. the time-independent white noise case $c(t)=1$ would make the the superpotential more nontrivial, and it will be dealt with in a future work.

\subsection{Symmetries and Ward identities}
To conclude this section, let us discuss the symmetries present in the {disordered electron gas} model and the  conserved currents associated with them. 

We will take a step back and look at the action~\eqref{eq:ac-with-B}. It describes the disorder problem for all Landau levels for nonzero $m$. It is symmetric under gauge transformations $(\delta A_0,\, \delta A_i)=(\partial_0\alpha, \, \partial_i \alpha)\,, \ (\delta\varphi,\, \delta \psi)=(i\alpha\varphi,  \, i\alpha \psi)$, and also under diffeomorphism transformations $x^i\to x'^i=x^i+\epsilon^i(x)$. These are associated with conservation of the current and the stress-tensor (see e.g. \cite{Geracie:2014nka}):
\begin{align}
   & \partial_0\rho+\partial_i J^i=0 \label{conseq-1}\\
    & \partial_jT^{ij}+\frac{1}{2}J^k\varepsilon_{ik}B=0 \label{conseq-2}\,.
\end{align}
Here $\rho, J^i, T^{ij}$ denote the charge density, gauge current and stress tensor respectively. They are defined as follows:
\be\label{sup-current}
\rho:=\frac{\delta S}{\delta A_{0}} \,, \ \ J^{i}:=\frac{\delta S}{\delta A_{i}}\,, \ \ T^{ij}:=2 \frac{\delta S}{\delta g_{ij}}\,.
\ee
Correlation functions of these currents are related to physical properties of the system. The conductivity tensor for example can be obtained from the two-point function $\langle J^i(x)J^j(y)\rangle$. 
The expectation is that the supersymmetry present in this model could (significantly) simplify computations of such correlation functions and give additional transport coefficients. However, as of now the equations above do not contain any information about the supersymmetry. 
It is difficult to incorporate that information in the analysis carried out so far. 
One would need to derive a supersymmetric version of the currents and conservation equations instead.

Let us now focus on the supersymmetry at the lowest Landau level. 
The chiral  SUSY action~\eqref{eq:susylll} is invariant under rotations in superspace which preserve the distance $z\zb+\theta\thetab$. These superrotations can be parametrized as $\delta z=ipz+\epsilon \theta , \delta\theta=iq\theta+\bar{\epsilon} z$ where $p, q$ are small and Grassmann even, while $\epsilon$ is Grassmann odd. { Setting $p,q=0$} and for small $\epsilon$, the corresponding field transformations $\{u(z),v(z)\}\to \{u'(z'),v'(z')\}=\{u(z')+\delta u(z'),v(z')+\delta v(z')\}$ that keep the action invariant can be written as:
\be\label{suprot}
\delta u(z)=-\frac{k}{\sqrt 2}\bar{\epsilon} z\, v(z)\,, \ \delta v(z)=-\frac{\sqrt 2}{k}\epsilon \, \partial_z u(z)\,.
\ee
Along-with the usual rotations these superrotations form the OSp$(2|2)$ supergroup. We also have symmetry under supertranslations, i.e. when $(z,\theta)\to (z',\theta')=(z+\delta_z,\theta+\delta_\theta)$ the action \eqref{eq:susylll} is invariant under the following superfield transformations:
\be\label{suptrans}
\delta U(z,\theta)=-\delta_z\partial_zU(z,\theta) -\delta_\theta \partial_\theta U (z,\theta)+\frac{k^2}{2}(\bar{\delta}_z z+\theta \bar{\delta}_\theta)\, U(z,\theta) \,,
\ee
and analogously for $U^{\dagger}(\zb,\thetab)$\,. The invariance under~\eqref{suptrans} is an extension of the usual `magnetic translational' invariance in the Landau level problem.  
We will therefore call~\eqref{suprot} and~\eqref{suptrans} \emph{magnetic-super-Poincare invariance}. 

The magnetic-super-Poincare symmetries are continuous symmetries and should be associated with conservation equations in superspace. In particular one should be able to derive conservation equations for a super-stress-tensor $\widetilde{T}^{ab}$ and -current  $\widetilde{J}^a$ multiplets (where $a$ can take the values $i,\theta,\thetab$) similar to the ones in the usual Parisi-Sourlas SUSY QFT (SQFT) of \cite{Parisi:1979ka}, which has exact super-Poincar\'{e} symmetry. In our case we should have a conservation law of the form \footnote{For exact  super-Poincare invariance T$^{d|2}\rtimes$OSp$(d|2)$, the super-stress-tensor  $T^{ab}(x^i,\theta_1,\theta_2)$ satisfies $\partial_a T^{ab}=0$, where $a=x^1,\cdots, x^d,\theta_1,\theta_2$.}
\be\label{expectedcons}
\partial_i \widetilde{T}^{ia}+\partial_{\theta}\widetilde{T}^{\theta a}+\partial_{\thetab}\widetilde{T}^{\thetab a}+\widetilde{J}^{a}=0\,.
\ee
In order to write down expressions for the supercurrents and super-stress-tensor one has to vary the Lagrangian~\eqref{eq:susylll} with respect to all its degrees of freedom. 
We leave this for section~\ref{sec:Wardid}, where we derive an improved version of these currents from a different starting point.  

Let us comment on how the usual currents $T_{ij}$ and $J^i$ are related to $\widetilde{T}^{ab}$ and $\widetilde{J}^a$. 
For the usual Parisi-Sourlas SQFT, the spatial components of the super-stress-tensor reduce to the usual stress-tensor on the $d$-dimensional manifold.
Similarly, the superconservation law~\eqref{expectedcons} should reduce to the conservation law for spatial diffeomorphism invariance~\eqref{conseq-2}, if we take $a=i$ and set the coordinates $\theta=\thetab=0$. Indeed the spatial part of these symmetry transformations (parameterized by $p$ and $\delta_z$ above) constitute the usual rotations and magnetic translations of our model. 
The question is now how $T^{ij}$ and $J^i$ relate to $\widetilde{T}^{ab}$ and $\widetilde{J}^{a}$ respectively.

This turns out to be a nontrivial question. We have mentioned the difficulty with deriving expressions for the currents after having fixed gauge and the metric. Instead, we should start from the gauge-invariant action~\eqref{eq:lagrangemult} coupled to a generic metric. Deriving expressions for $T^{ij}$ and $J^i$ is now straightforward, using the definitions~\eqref{sup-current}.
However, they will not only depend on the superfields  but on the auxiliary fields $\chi_1, \chi_1^\dagger$ as well. 
These auxiliary fields in general break supersymmetry, and do not appear in the SUSY action~\eqref{eq:susylll}. Therefore they will not appear in $\widetilde{T}^{ab}, \widetilde{J}^a$, and we are unable to simply reduce the supercurrent and super-stress-tensor to $T^{ij}$ and $J^i$ on the $d$-dimensional manifold, as suggested above.

If we want to make supersymmetry manifest in the expressions for the currents and stress-tensor, and in the corresponding conservation equations, we will need to rewrite~\eqref{eq:susylll} in a different form, starting from a gauge-invariant theory coupled to a metric. This will be the topic of the next section.

\section{Parisi-Sourlas SUSY in external magnetic field}\label{sec:gensusy}

The chiral SUSY theory that emerges in the LLL of the disordered electron gas model is similar to the emergent Parisi-Sourlas SCFT in random field models (see sec.~\ref{sec:intro}) as both have an underlying OSp$(d|2)$ supersymmetry of superrotations. To compare the two SUSY theories let us first recall from section \ref{sec:intro} our definition of a PS theory:
\be\label{PS-susy-1}
S_{\text{PS}}[\Phi]=\int \text{d}^{2|2}y\, \Big[ \frac{1}{2}|\partial_a\Phi|^2+V(|\Phi|) \Big]\,,
\ee
The  superfield $\Phi$ can be expanded in $\theta_1,\theta_2$. However, for later convenience, we will write this expansion in complex coordinates $\theta, \thetab$, that we defined in~\eqref{complex-Grassmann}, as follows:
\begin{align}
    &\Phi(y)=\Phi_0(x)+\theta \Phi_{\theta}(x)+\thetab \Phi_{\thetab}(x)+ \theta\thetab \Phi_{\theta\thetab}(x)\,, \label{Phisuperfield} \\
    &\Phi^{\dagger}(y)=\Phi^{\dagger}_0(x)+ \Phi^{\dagger}_{\theta}(x) \thetab + \Phi^{\dagger}_{\thetab}(x) \theta+ \theta\thetab \Phi^{\dagger}_{\theta\thetab}(x)\,.
\end{align}
All components of this decomposition are complex valued, and $\Phi_{\theta}, \Phi_{\thetab}$ are Grassmann odd. The action has a T$^{d|2}\rtimes$OSP$(d|2)$ super-Poincar\'{e} invariance.
The superspace index $a$ takes values $(1,\cdots ,d, d+1, d+2)$. Accordingly $y^a=\{x_1,\cdots,x_d,\theta_1,\theta_2\}$, where $x_i$ and $\theta_1,\theta_2$ are (real) bosonic and fermionic coordinates respectively.
Finally, the superspace volume element is given by $\text{d}^{d|2}y:=\text{d}^dx\text{d}\theta\text{d}\thetab$, and recall that our conventions are such that the fermionic coordinates obey  the Grassmann integral~\eqref{complex-Grassmann-int} i.e. $\int \td \theta\td \thetab\, \thetab\theta=1$. \\ \\
The usual PS theories have a fixed point with an extended superconformal invariance that also emerges at the fixed point of RF models (see the discussion around Fig. \ref{FigRFmodel} in section \ref{sec:intro}). Now compare this with the LLL limit of the { disordered electron gas} model which is an emergent chiral SUSY theory~\eqref{eq:susylll}  with magnetic super-Poincar\'e invariance. It is natural to wonder if there exists a `Parisi-Sourlas-like' theory $S_{\text{PS}}'$ that is coupled to an external magnetic field-like parameter $\widetilde{B}$, and for large $\widetilde{B}$ it reduces to~\eqref{eq:susylll}, i.e. 
\be\label{lllanalogy}
S'_{\text{PS}}[\Phi(z,\zb,\theta,\thetab)] \, \stackrel{\widetilde{B} \gg 1}{\longrightarrow} \, S[U(z,\theta)] \,.
\ee
This resembles a SUSY version of the non-disordered electron gas problem that reduces to a theory of holomorphic and anti-holomorphic fields in the LLL limit (similar to~\eqref{eq:LLL-th} but without the fields $v,v^\dagger$). If such a theory exists, there are two possible scenarios:
\begin{enumerate}
    \item This $S'_{\text{PS}}$ model reduces to the chiral SUSY theory~\eqref{eq:susylll}  for large $\widetilde{B}$, but is not connected to the disordered  problem for finite $\widetilde{B}$.
    \item The disordered electron problem can be mapped to $S'_{\text{PS}}$ for any $\widetilde{B}$. 
\end{enumerate}
The first case is similar to the situation of RF models, where the Parisi-Sourlas SUSY is absent along the RG flow but only emerges at the fixed point (see Fig. \ref{FigRFmodel}). 
At first sight this seems a likely scenario for our disordered model as well. Indeed the dimensional reduction which appears as a result of the emergent SUSY, is only observed at the LLL and not for finite $B$ (or $m$). 
However, a modification of case 2 can still be possible. For finite $\widetilde{B}$, even though SUSY might be broken there could be a description in terms of SUSY fields and coordinates. This will imply a SUSY-like formulation of the whole problem, which would be interesting.

\subsection{A magnetic Parisi-Sourlas theory}\label{sec:proposal}
In this subsection we will construct a modified  Parisi-Sourlas theory as discussed above. An obvious question is what the new $\widetilde{B}$ parameter mentioned above~\eqref{lllanalogy} could be and how to couple it. Recall from the global OSp model~\eqref{scalar-EFT} that the role of taking large $B$ was played by setting $m\to 0$. 
Comparing with that theory we start with a nonrelativistic version of the SUSY action~\eqref{PS-susy-1}, and introduce two external parameters $A_a(y)$ and $(\varepsilon^{2|2})^{ab}$, defined below, which modify the action in the following way:
\be\label{eq:susy-gen}
S[\Phi]=\int \text{d}t\,\text{d}^{2|2}y\, \Big[{i}\Phi^{\dagger}{D_0} \Phi- \frac{1}{2m}((g^{(2|2)})^{ab}+i(\varepsilon^{(2|2)})^{ab}) (\widetilde{D}_a \Phi^\dagger) (D_b \Phi)+ V(\Phi) \Big]\,,
\ee
where $V(\Phi)$ can be an arbitrary interaction symmetric in $\Phi, \Phi^{\dagger}$ and is not limited to the quartic interaction $V(\Phi) = \lambda(\Phi^\dagger \Phi)^2$ used in the previous section. 

We will call $S[\Phi]$ a magnetic Parisi-Sourlas theory. Its Lagrangian is defined in superspace $\mathbb{R}^{1,2|2}$ parametrized by $(t,y^a)$. We treat the time-coordinate $t$ separately, as the theory is non-relativistic. 
The indices $a,b$ take values $(1,2,3,4)$ and $y^a=\{x_1,x_2,\theta_1,\theta_2\}$. For convenience, we will often denote the bosonic components with $i,j$ indices, i.e. $y^i=\{x_1,x_2\}$ and fermionic components with $p,q$ indices i.e. $y^p=\{\theta_1,\theta_2\}$. The  definition of superfield $\Phi(t,y)$ is identical to that of~\eqref{Phisuperfield}.\\ \\
We have also introduced  super-covariant derivatives $D_0=\partial_0-iA_{0}(x,\theta)$, $D_a=\partial_a-iA_{a}(x,\theta)$, and $\widetilde{D}_a = \partial_a + i A_a$. Here $A_a$ denotes a four-component background field. The theory has a generalized gauge symmetry under
\be
A_0\to A_0+\partial_0 \alpha(t,y)\,, \ A_a\to A_a+\partial_a \alpha(t,y)\,,  \ \Phi\to e^{i\alpha(t,y)}\Phi\,.,
\ee
where $\alpha(t,y)$ is an OSp$(2|2)$ real scalar function. \\ \\
The OSp$(2|2)$ metric $g^{(2|2) ab}$ is given by
\be
g^{(2|2) ab}:=\left(\begin{array}{cc}
 \eta & \ 0 \\
 0 & \ i\,\varepsilon  \end{array}\right)\,, \ \eta^{ij}:= \left(\begin{array}{cc}
 1 & \ 0 \\
 0 & \ 1 \\\end{array}\right)\,, \ \varepsilon^{pq}:=\left(\begin{array}{cc}
 0 & \ 1 \\
 -1 & \ 0 \\\end{array}\right)\,, \quad g^{(2|2) ab} = g^{(2|2)}_{ab}\:.
\ee
We have also introduced a graded antisymmetric tensor $\varepsilon^{(2|2)}$, which is a block matrix of the form
\be\label{susy-varep}
\varepsilon^{(2|2) ab} :=\left(\begin{array}{cc}
\varepsilon & \ 0 \\
 0 & \ i\, \eta  \end{array}\right)\,.
\ee
The reader should be warned that $\varepsilon^{(2|2)}_{ab}$ is not an  invariant antisymmetric tensor under all   OSp$(2|2)$ superroations.  It is simply a constant graded antisymmetric tensor present in the definition of $S[\Phi]$. However we can use it to associate a `super-magnetic field' $\widetilde{B}=\veps^{(2|2) ab} \partial_a A_b$ to the external field $A_a$. But $\widetilde{B}$ is not a tunable parameter yet since $A_a$ is still unspecified. \\ \\
Notice that when $A_a=0$ the theory gets no contribution from the $\varepsilon^{(2|2)}$ term. This theory then is a pure PS theory of the form~\eqref{PS-susy-1}, but with an additional time derivative, i.e. it is symmetric under supertranslations and superrotations that preserve the distance $y^2=g^{(2|2)}_{ab}y^ay^b=x_1^2+x_2^2+2\,i\,\theta_1\theta_2$\,.
The generators of supertranslations and superrotations  satisfy a set of graded commutation relations which form the super-Poincar\'{e} algebra, and are straightforward generalizations of usual Poincar\'{e} generators. We review it compactly in appendix \ref{app:spa}. 
 \\ \\
For arbitrary, non-zero $A_a$ the super-Poincar\'{e} invariance is broken and~\eqref{eq:susy-gen} is written as follows:
\begin{align}\label{eq:action-new}
S[\Phi]&=\int \td^2x \td^2\theta\, \Big[i \Phi^{\dagger}{D_0} \Phi- \frac{1}{2m}({(g^{(2|2)}})^{ab}+i {(\varepsilon^{(2|2)}})^{ab})(\widetilde{D}_a \Phi^\dagger )(D_b \Phi)+V(\Phi) \Big]\nonumber\\
&=\int \td^2x \td^2\theta\, \Big[i \Phi^{\dagger}{D_0} \Phi- \frac{1}{2m}(\eta^{ij}+ i\varepsilon^{ij})(\widetilde{D}_i \Phi^{\dagger}) D_j \Phi + \frac{1}{2m}(\eta^{pq}-i\varepsilon^{pq})(\widetilde{D}_p \Phi^{\dagger}) D_q \Phi+V(\Phi) \Big]\,,
\end{align}
We would like to mimic the approach to LLL of the non-SUSY electron gas case by taking the $m\to 0$ limit.
The above structure lets us repeat our analysis in subsection \ref{sec:proj-LLL}. We repeat the step from \eqref{eq:step-intro-X} for both the pair of indices $(ij)$ and $(pq)$ and write the covariant derivatives in~\eqref{eq:action-new} as: 
\be\label{eq:step-intro-X-susy}
(\eta^{ij}+ i \epsilon^{ij})(\widetilde{D}_i \Phi^{\dagger}) D_j \Phi-(\eta^{pq}- i \epsilon^{pq})(\widetilde{D}_p \Phi^{\dagger}) D_q \Phi \ = \  4 (e^i \widetilde{D}_i \Phi^{\dagger}) (\bar{e}^j D_j \Phi) - 4 (e^p \widetilde{D}_p \Phi^{\dagger}) (\bar{e}^q D_q \Phi)\,
\ee
where we have introduced another set of complex Zweibeine for the $p,q$ indices in a similar way as for the $i,j$ indices in~\eqref{eq:zweibeinedef1}, but with opposite signs:
\be
e^p = \frac{1}{2} (\delta^{3 p} + i \delta^{4 p})\:, \quad \bar{e}^p = \frac{1}{2} (\delta^{3 p} - i \delta^{4 p})\:, \quad e^p e_p = \bar{e}^p \bar{e}_p = 0\:, \quad e^p \bar{e}_p = - \bar{e}^p e_p = \frac{1}{2}\:.
\ee
These allow us to write the fermionic parts of the metric and antisymmetric tensor as follows:
\be
g^{(2|2) pq} = 2 \, (\bar{e}^p e^q - e^p \bar{e}^q)\:, \quad \varepsilon^{(2|2) pq} = 2 i \, (e^p \bar{e}^q + \bar{e}^p e^q)\:,
\ee
and again the identity~\eqref{eq:step-intro-X-susy} follows immediately.
The terms of the form $e^aD_a\Phi$ appear quadratically but with a singularity at $m=0$. So we use the Hubbard-Stratonovich transformation as in~\eqref{eq:aux-fields}, and introduce two auxiliary superfields $X_1$ and $X_2$ to rewrite the quadratic terms: 
\begin{align}\label{susy-X1X2}
S[\Phi,X_1,X_2]=\int \td^2x \td^2\theta\, &\Big[i \Phi^{\dagger}{D_0} \Phi -  X_1^\dagger(\bar{e}^iD_i\Phi) - (e^i \widetilde{D}_i\Phi^\dagger) X_1 + X_2^\dagger(\bar{e}^p D_p\Phi) + (e^p \widetilde{D}_p\Phi^\dagger) X_2\nonumber\\
& + \frac{m}{2} (X_1^\dagger X_1 - X_2^\dagger X_2)+V(\Phi) \Big]\,.
\end{align}
Once again this results in a smooth $m\to 0$ limit, where the auxiliary superfields become Lagrange multipliers. 
At this point let us fix the external SUSY gauge field to be of the form:
\be\label{Achoice}
A_{a}=
 \left(\begin{array}{rr}
 -\frac{B}{2}\, y&   \\
 \frac{B}{2}\, x &  \\
- \frac{i\, B'}{2} \, \theta_1 &  \\   - \frac{i\,B'}{2}\, \theta_2 &  \end{array}\right)\,.
\ee
Therefore the $\widetilde{B}$ parameter desired in the beginning of this section is essentially a combination of the two independent parameters $B,B'$; and from analogy with our sec \ref{sec:proj-LLL} arguments  the largeness of both are enforced by small $m$. Now, with $m=0$, 
varying \eqref{susy-X1X2} with respect to $X_1^\dagger$ and $X_2^\dagger$ gives respectively: 
\begin{align}
    D_{\zb} \, \Phi &=0\, \ \ \text{with} \ \  D_{\zb}:=\bar{e}^iD_i=\partial_{\zb} + \frac{B}{4} z\,, \label{eq:lll-sp} \\
    D_{\thetab} \, \Phi&=0\, \  \ \text{with} \ \  D_{\thetab}:= \bar{e}^p D_p=\partial_{\thetab} - \frac{B'}{4}\theta \label{eq:lll-theta}\,.
\end{align}
We have analgous equations for the complex conjugates. \\ \\ 
We will call \eqref{eq:lll-sp} and \eqref{eq:lll-theta} bosonic and fermionic LLL conditions respectively, and them together as `SUSY LLL' conditions. 
The first condition in~\eqref{eq:lll-sp} and its complex conjugate are the usual lowest Landau level constraints on the superfield $\Phi$ and $\Phi^\dagger$ respectively as found before in~\eqref{eq:lll-sp-def}. This implies/enforces that each component of $\Phi$ in the decomposition~\eqref{Phisuperfield} must be of the form of the LLL wave-function~\eqref{lll-sol}. The second condition~\eqref{eq:lll-theta} is new. It has a solution given by
\begin{align}\label{thetalll-sol}
\Phi_{\thetab} &= 0\,, \ \ \Phi_{\theta\thetab} = -\frac{B'}{4} \Phi_0\,.
\end{align}
The form \eqref{Achoice} was picked specifically to have the desired form of SUSY LLL conditions. 
If we take $B'=B$ then we obtain the general solution to the conditions
\begin{align}\label{sLLLsol}
\Phi(z,\zb,\theta,\thetab)&=e^{-\frac{B^2}{4}(|z|^2+\theta\thetab)}U(z,\theta)\,, \ \ U(z,\theta)=u(z)+\sqrt{\frac{B}{2}}\, \theta\, v(z)\,.
\end{align}
A similar equation for $\Phi^\dagger$ can be found by complex conjugation.
Choosing $A_0=A_0(x)$ and $V(\Phi) = \lambda(\Phi^\dagger \Phi)^2$, we find that in the $m\to 0$ limit the action~\eqref{susy-X1X2} reduces to 
\be\label{result1}
\boxed{S[\Phi,X_1,X_2] \ \stackrel{m\to 0}{\longrightarrow} \ S_{\text{chiral}}[\Phi]\, =\, \int \text{d}t\,\text{d}z \text{d}\bar{z}\, \text{d}\theta \text{d}\bar{\theta}  \, \big(i \Phi^{\dagger}D_0 \Phi+\lambda(\Phi^\dagger \Phi)^2 \big)\,,}
\ee
where $\Phi,\Phi^{\dagger}$ are restricted to the SUSY-LLL solutions \eqref{sLLLsol} in the path integral. In other words, $S_{\text{chiral}}[\Phi]$ is identical to the chiral SUSY theory $S[U]$ in \eqref{eq:susylll} that was derived from the disordered theory using the replica trick. 

\subsection{Special gauge choices}
\label{sec:specialgauge}
Enforcing the new SUSY LLL conditions has enabled us to make a connection between this model and our original disordered theory in the $m \to 0$ limit, and with the choice of gauge field specified in~\eqref{Achoice}. 
We will now explore two alternative, special cases that correspond to other particular choices of $A_a$.

\paragraph{Case I:} In this case we only fix the $A_p$ components of the gauge field, meaning we choose $A_a$ to be 
\be\label{gauge1}
A_{a}=
 \left(\begin{array}{cc}
 A_1(x,y)&   \\
  A_2(x,y) &  \\
-\frac{i\, B'}{2} \, \theta_1 &  \\   -\frac{i\,B'}{2}\, \theta_2 &  \end{array}\right)\,.
\ee
We may still project onto the SUSY LLL conditions by taking $m\to 0$, but in the condition~\eqref{eq:lll-sp} one cannot fix the form of $D_{\bar{z}}=\bar{e}^iD_i$ to be $\partial_{\zb} + \frac{B}{4} z$.  The fermionic LLL 
condition~\eqref{eq:lll-theta} however is unchanged, and we may impose it exactly. In the action~\eqref{susy-X1X2} this means dropping the $X_2$ and $X_2^\dagger$ terms and requiring~\eqref{thetalll-sol} in the path integral.
Let us emphasize that now we do not fix the forms of $\Phi_0,\Phi_\theta$ in terms of holomorphic fields as shown in~\eqref{lll-sol}.
We then get the action~\eqref{eq:lagrangemult} with $m=0$, which is our original replica theory at $n\to 0$ restricted to the LLL: \footnote{While both $\Phi$ and $X_1$ begin with four components, with the fermionic LLL conditions in place there are two independent nonzero components of $\Phi$. The corresponding two components of $X_1$ can be identified with $\vec\chi=\{\chi_1,\chi_2\}$. The other two components of $X_1$ decouple from the theory.}
\be\label{result2}
\boxed{S[\Phi,X_1,X_2]\Big|_{\{m=0,A_p=\varepsilon^{pq}y_q\}}\  \to \ S[\vec \phi,\vec \chi]\Big|_{m=0}}
\ee
Note that we do not get the theory of~\eqref{eq:susylll}, which is a SUSY theory for the holomorphic fields $u(z)$ and $v(z)$. This shows us that there is a way to relate the disorder averaged theory at the LLL to a SUSY theory, namely~\eqref{susy-X1X2}, without fixing the bosonic components of the gauge field. This SUSY theory is different from the one which emerges at the LLL for a particular gauge.

\paragraph{Case II: \label{higherLL}} Another interesting choice for $A_a$ is setting $B':=\alpha\, B\gg B$ in \eqref{Achoice}:
\be\label{gauge-choice}
A_{a}=
 \frac{B}{2}\left(\begin{array}{cc}
  -y&   \\
  \, x&   \\
i \, \alpha^2\, \theta_1&   \\   i\,\alpha^2\, \theta_2&   \end{array}\right)\,,\hspace{1cm} \alpha \gg 1\,.
\ee
This corresponds to a large `fermionic' part of the super-magnetic field. 
Unlike the analysis so far, let us not consider $m$ to be small. We show below that there is still a singularity but only in the fermionic covariant derivatives of $S[\Phi]$. So let us focus on that term from~\eqref{eq:action-new} separately and write $m=m'\alpha^2$:
\be\label{intstep}
\frac{1}{2m}(\eta^{pq}-i\varepsilon^{pq})(\widetilde{D}_p \Phi^{\dagger}) D_q \Phi= \frac{1}{2m'}(\eta^{pq}-i\varepsilon^{pq})(\frac{1}{\alpha}\widetilde{D}_p \Phi^{\dagger} )\frac{1}{\alpha}D_q \Phi\,.
\ee
To keep $m$ finite we must have $m'\ll 1$ if $\alpha \gg 1$.
Now let us introduce a new set of rescaled Grassmann coordinates $(\theta_1,\theta_2)\to (\hat{\theta}_1,\hat{\theta}_2)=(\alpha\,\theta_1,\alpha\,\theta_2)$. In the new coordinates $\hat{y}:=\{x,y,\hat{\theta}_1,\hat{\theta}_2\}$ we define new fermionic covariant derivatives
\begin{align}
    \widehat{D}_p:=\widehat{\partial}_p-i\,\hat{A}_p\:, \quad \widehat{\partial}_p=\alpha^{-1}\partial_{p}\:, \quad \hat{A}_p = \frac{B}{2}\,(i\, \hat{\theta}_1,i\, \hat{\theta}_2) = \alpha^{-1} A_p\:,
\end{align}
such that $D_p = \alpha \widehat{D}_p$ and similarly for $\widetilde{D}_p$. Then all $\alpha$'s in the r.h.s. of~\eqref{intstep} are removed. 
The superfields also rescale: $\Phi(y)=\hat{\Phi}(\hat{y})$
Then~\eqref{intstep} becomes
\be
\frac{1}{2m}(\eta^{pq}-i\varepsilon^{pq})(\widetilde{D}_p \Phi^{\dagger}) D_q \Phi= \frac{1}{2m'}(\eta^{pq}-i\varepsilon^{pq})(\widehat{\widetilde{D}}_p \hat{\Phi}^{\dagger}) \widehat{D}_q \hat{\Phi}\:.
\ee
The matrices $\eta^{pq},\varepsilon^{pq}$ are constant and unchanged. Using the identity~\eqref{eq:step-intro-X-susy}, the full action can be rewritten similar to~\eqref{susy-X1X2}, with the only replacement $m |X_2|^2\ \to \ m'|\hat{X}_2|^2$\,. 

At this stage the only presence of $\alpha$ in the action is through the measure $\td^2x\,\td^2\theta=\alpha^{-2}\td^2x\,\td^2\hat{\theta}$. Since this is an overall factor, we ignore it and remove all hats from the coordinates and fields for simplicity, resulting in:
\begin{align}
S[\Phi,X_1,X_2]\propto \int \td^2x \td^2\theta\, &\Big[i \Phi^{\dagger}{D_0} \Phi -  X_1^\dagger(\bar{e}^iD_i\Phi) - (e^i \widetilde{D}_i\Phi^\dagger) X_1 + X_2^\dagger(e^p D_p\Phi) + (\bar{e}^p \widetilde{D}_p\Phi^\dagger) X_2\nonumber\\
& + \frac{m}{2} X_1^\dagger X_1 - \frac{m'}{2}X_2^\dagger X_2)+V(\Phi) \Big]\,.
\end{align}
Now in the small $m'$ limit we can eliminate $X_2$ (but not $X_1$), and again obtain the fermionic LLL condition~\eqref{eq:lll-theta}. This time, restricting $\Phi$ to the solution~\eqref{thetalll-sol}, we arrive at 
\be\label{result3}
\boxed{S[\Phi,X_1,X_2]\Big|_{B'\gg B}\to S[\vec \phi]\Big|_{m\neq 0}}
\ee
i.e. a connection between our SUSY model and the disordered electron gas without imposing the $m \to 0$ limit! 
This particular case motivates our scenario 2, described at the beginning of this section. 

\subsection{Dimensional reduction}\label{dimredsec}

An important consequence of the magnetic Parisi-Sourlas theory~\eqref{eq:susy-gen}  is dimensional reduction of a sector of correlation functions in the SUSY LLL limit $m\to 0$. We will establish this in this section. 
This dimensional reduction was originally established for the specific case of a two-point function in \cite{BREZIN198424} with a more direct approach. We briefly review their findings below. 

The main quantity of interest in \cite{BREZIN198424} is 
the density of states per unit area $\mathcal{D}(E)$ at an energy $E$ \cite{Wegner:2016ahw}, which is related to the disorder averaged two-point function in the disordered electron gas model at the coincident limit $y_1=y_2$ as follows:
\be
\mathcal{D}(E)=\text{Im }\mathcal{F}_E\big[\,\overline{\langle \phid(t,z,\zb)\phi(t,z,\zb)\rangle} \,\big]=\text{Im }\mathcal{F}_E\big[\,\langle \vf^\dagger(t,z,\zb)\vf(t,z,\zb)\rangle \,\big]\,.
\ee
Here $\mathcal{F}_E[f(t)]$ denotes the Fourier transform of a function $f(t)$. The second equality is a consequence of the equivalence \eqref{avg-repl-2} and the r.h.s. of the second equality is computed in global OSp theory of~\eqref{eq:ac-with-B}.

Now $\vf$ is the $\theta,\thetab=0$ component of the superfield $\Phi$, under the SUSY LLL restrictions~\eqref{eq:lll-sp} and~\eqref{eq:lll-theta}. Therefore $\mathcal{D}(E)$ can be related to the two-point function $\Phid$-$\Phi$, which has the general form:
\begin{align}
\langle\Phid(t,y_1)\Phi(t,y_2)\rangle&=\langle U^\dagger(\zb_1,\thetab_1)U(z_2,\theta_2) \exp\big({-\frac {k^2}4 (z_1\zb_1+\theta_1\thetab_1)-\frac {k^2}4 (z_2\zb_2+\theta_2\thetab_2)}\big) \label{2ptsusy} \\
&=  C_t\,e^{\frac{k^2}{2}(z_2\zb_1+\theta_2\thetab_1)}\exp\big({-\frac {k^2}4 (z_1\zb_1+\theta_1\thetab_1)-\frac {k^2}4 (z_2\zb_2+\theta_2\thetab_2)}\big)\label{PhiPhi-gen}\,.
\end{align}
The first equality simply follows from the SUSY LLL functional form of $\Phi$ given in~\eqref{sLLLsol}. To write the second equality we used the fact that the two-point function $\langle U^\dagger U \rangle$ should be invariant under the { magnetic} super-translations~\eqref{suptrans}. In the coincident limit $y_1=y_2$ the $z,\theta$ dependencies drop from~\eqref{PhiPhi-gen}. The quantity $C_t$ is simply $\langle \Phid(t,y_1=0)\Phi(t,y_2=0)\rangle$. This two-point function can be directly computed perturbatively in the chiral SUSY theory~\eqref{eq:susylll}  using Feynman diagrams and it gives 
\be\label{Grossetalresult}
C_t=\frac{\int \mathcal{D}\phi(t)\, \phi^{\dagger}(t)\phi(t)\, e^{-S_{(0)}[\phi]}}{\int \mathcal{D}\phi(t)\, e^{-S_{(0)}[\phi]}}
\ee
where
\be\label{dimredac}
S_{(0)}[\phi]=\int \td t \, \Big[{i}\phi^{\dagger}D_0 \phi+V(\phi) \Big]\,
\ee
is a $0+1$ dimensional theory. The interaction $V(\phi)$ has the same form as $V(\Phi)$ up to a factor of $\pi$.\footnote{This $\pi$ can be set to 1 if choose the $\theta,\thetab$ integral measure in \eqref{complex-Grassmann-int} differently, see footnote \ref{pifootnote}.} Eq.~\eqref{Grossetalresult} implies that we have essentially reduced the 2-point function computation to a quantum mechanics problem. 

In \cite{BREZIN198424} it was shown that each internal vertex of the Feynman diagrams has a cancellation between the bosonic $(z,\zb)$ and fermionic $(\theta,\thetab)$ coordinates that results in the above  result. We point out that in \cite{BREZIN198424} the starting point is a 2d disordered theory (no time axis) that reduces to a 0d action (whose `path integral' is an ordinary integral). Since we started with a model in $2+1$d, the dimensionally reduced theory has a $t$ direction. \\ \\
We will show that one can make a stronger statement of dimensional reduction for the
magnetic Parisi-Sourlas theory~\eqref{eq:susy-gen}. For an arbitrary correlation function of $\Phid, \Phi$, one has 
\be\label{dimred-gen}
\boxed{\langle\Phid(x_1)\cdots\Phid(x_m)\Phi(x_{m+1})\cdots\Phi(x_n)\rangle\Big|_{S[\Phi]}\stackrel{m\to 0}{=} \langle \phid(x_1)\cdots\phid(x_m)\phi(x_{m+1})\cdots\phi(x_n) \rangle\Big|_{S_{(0)}}\,,}
\ee
where the correlator on the right is computed using~\eqref{dimredac}.

To prove~\eqref{dimred-gen} let us start with the path integral representation of the l.h.s. and replace the action with $S[\Phi,X_1,X_2]$ in~\eqref{susy-X1X2}, which does not have any singularity at $m\to 0$. 
Now the full action $S[\Phi,X_1,X_2]$ can be written as a sum of two components as follows:
\begin{align}
&S[\Phi,X_1,X_2]= S_{\text{PS}}[\Phi] + S_{(1)}[\Phi,X_1,X_2]\,,\label{spn-PS}\\
&S_{\text{PS}}[\Phi]:=\int \text{d}t\,\text{d}^{2|2}y\, \Big[i\Phi^{\dagger}D_0 \Phi+V(\Phi) \Big]\label{action-PS}\,,\\
&S_{(1)}[X_1,X_2,\Phi]:= \big( - X_1^\dagger(\bar{e}^iD_i\Phi) +  X_2^\dagger(\bar{e}^pD_p\Phi)\big)+\text{c.c.}  + \frac{m}{2} (X_1^\dagger X_1- X_2^\dagger X_2) \label{S-extra}\,.
\end{align}
The point of separating out the action is that $S_{\text{PS}}$ is a pure Parisi-Sourlas SUSY of the form \eqref{PS-susy-1}, which we introduced in the introduction. Any correlation function of the type $\langle \Phi(t_1,y_1)\cdots\Phi(t_n,y_n)\rangle$ in $S_{\text{PS}}$ is dimensionally reduced from $2+1$ dimensions to $0+1$ dimension when all $y_i=0$, and is given by the action~\eqref{dimredac}. This can be understood in perturbation theory by considering all constituent Feynman diagrams that describe this correlation function \cite{PSproof2,CARDY1985123}. The general $d\to d-2$ dimensional reduction of an arbitrary diagram is reviewed in App. A of \cite{Kaviraj:2019tbg}. The main element of the proof is the following identity applied to each vertex of the diagram:
\be
\int\, \td^{2|2}y\, f(\hat x,y^2)=\pi\, f(\hat x,0)\,.
\ee
Here $f$ is a general function which may be identified with a product of propagators. Here $\hat x$ are coordinates of the $d-2$ hyperplane $y=0$, and $\hat x=t$ for our setup. The identity shows that the internal lines of a generic diagram get restricted to the $y=0$ manifold. The resulting amputated diagram is the same as a diagram constructed from~\eqref{dimredac}.

In our case the full theory is given by~\eqref{spn-PS} and it has the part $S_{(1)}$ that breaks the pure Parisi-Sourlas supersymmetry due to the $D_i$ terms. Still the dimensional reduction~\eqref{dimred-gen} holds because these SUSY-offending terms decouple from correlation functions of $\Phid$ and $\Phi$ as $m\to 0$. To see this, consider any generic Feynman diagram that contributes to such a correlator, i.e. the l.h.s. of~\eqref{dimred-gen}. The vertices come from the quartic interaction term in the $S_{\text{PS}}[\Phi]$ part of the action. Meanwhile all internal and external lines are $\Phid$-$\Phi$ propagators that may be written as
\be
G_{\Phi}(t,y)=G^{\text{PS}}_{\Phi}(t,y)+G^{1}_{\Phi}(t,y)
\ee
Here $G^{\text{PS}}_\Phi=(iD_0)^{-1}$ is obtained by inverting the quadratic term in $S_{\text{PS}}$. The other part $G^{1}_{\Phi}$ comes from the action $S_{(1)}[X_1,X_2,\Phi]$, which is fully quadratic. For $m\to 0$ it is of the form:
\be\label{quad-S1}
S_{(1)}[X_1,X_2,\Phi] \ \stackrel{m\to 0}{=} \ \left( \begin{array}{c}
\Phi \\ X_1 \\ X_2\end{array} \right)^\dagger
\left(\begin{array}{ccc}
0 & \ \ast& \ \ast  \\
 \ast & \ 0  & \ 0 \\ 
 \ast & 0 & 0\end{array}\right)  \left( \begin{array}{c}
\Phi \\ X_1 \\ X_2\end{array} \right)  \,,
\ee
Here $\ast$ indicates nonzero entries. By inverting the matrix it is easy to see that the $S_{(1)}$ part does not have a $\Phi^\dagger$-$\Phi$ propagator. The off-diagonal propagators such as $X_1^\dagger$-$\Phi$ or $X_2^\dagger$-$\Phi$ exist, but they cannot enter the diagram under consideration since all external fields as well as all vertices are built from $\Phi^\dagger, \Phi$ only.

This completes the proof of the dimensional reduction of correlators shown in~\eqref{dimred-gen}, which is a generalization of the two-point case shown in \cite{BREZIN198424}.  We have not found a similar derivation for such a general set of observables in the literature. 
Note that dimensional reduction will fail if one of the external points have an $X_1, X_2$ type of field, or if $m\neq 0$. We discuss in section \ref{sec:outlook} how some simplification could still happen in these cases. Notice that this subsection does not refer to any gauge choice. This means if we connect to disorder observables by picking $A_a$ to be of the form~\eqref{gauge1}, we will see dimensional reduction without gauge fixing $A_i$.

\section{Supersymmetric Ward identities}\label{sec:Wardid}
Having found a gauge-invariant action~\eqref{eq:susy-gen} that reduces to the LLL SUSY theory of section~\ref{sec:model}, we now return to the derivation of supercurrents and their conservation equations. We will show that supersymmetry implies 
these super-currents must satisfy a set of super-Ward identities.
\subsection{Supercurrents and super-conservation conditions}\label{supercurrents}
Our starting point for the derivation will be the action~\eqref{eq:susy-gen}, containing the explicit dependence on the background metric. The term proportional to the antisymmetric tensor $\varepsilon^{(2|2)}_{ab}$ can be rewritten in terms of a super-magnetic field $\widetilde{B}$, defined as 
\be
    \widetilde{B} = B + B' = \varepsilon^{(2|2) ab} \partial_a A_b\:.
\ee
In this section we will denote $g_{ab}^{(2|2)}$ as $g_{ab}$ for brevity. The action then takes the form 
\be\label{eq:susy-gen-met}
S[g,A,\Phi]=\int \text{d}t\,\text{d}^{2|2}y\,\sqrt{|g|} \Big[i \Phi^{\dagger} D_0 \Phi- \frac{g^{ab}}{2m}(\widetilde{D}_a \Phi^{\dagger})(D_b \Phi)+\frac{\widetilde{B}}{2m}|\Phi|^2+V(\Phi) \Big]\,.
\ee
The action $S[g,A,\Phi]$ now dependends on the external fields $A_0, A_a$ and $g_{ab}$. We define the supercurrents $\rho$, $J^{a}$ and $T^{ab}$ by respectively varying the action with respect to those external fields:
\be\label{var-ac}
\delta S=\int \text{d}t\,\text{d}^{2|2}y\big(\frac{1}{2} T^{ab}\delta g_{ab}+\rho \delta A_{0} +J^{a}\delta A_{a}\big)\,.
\ee
In principle we should also account for the variation with respect to $\widetilde{B}$, since it is an external parameter which is related to $A_a$ through a tensor parameter $\epsilon_{ab}$. However, we consider transformations that keep $\widetilde{B}$ constant. 
The variations with respect to $\Phi, \Phi^{\dagger}$ give the equations of motion. Notice that since~\eqref{eq:susy-gen-met} has a singularity for $m\to 0$, these definitions of the currents need to be slightly modified before going to the LLL theory. We will come back to this point in subsection \ref{sec:supcurr}. \\ \\
Consider now the symmetry of the action under the supersymmetric gauge transformations $\delta A_0=\partial_0 \alpha$, $\delta A_a=\partial_a \alpha$,  $\delta \Phi=i\alpha\Phi$. This implies the following identity
\begin{align}
&0=\int \text{d}t\,\text{d}^{2|2}y \, \Big[\rho \, \partial_0 \alpha(t,y)+ J^{a}\, \partial_{a}\alpha(t,y) \Big]\,\nonumber \\
&\implies \, \boxed{\partial_0\rho+\partial^{a}J_{a}=0}\label{cons1}\,.
\end{align}
We also have a symmetry under diffeomorhism $y^{a}\to y^{a}+\epsilon^{a}(y)$ under which the fields transform as:
\begin{align}\label{diff-trans}
&\delta\Phi=\epsilon^{a}\partial_{a}\Phi\nonumber\,,\\
&\delta A_{a}=\epsilon^{b}\partial_{b}A_{a}+\partial_{a}\epsilon^{b}A_{b}\,,\nonumber\\
&\delta g_{ab}=\epsilon^{c}\partial_{c}g_{ab}+\partial_{b}\epsilon^{c}g_{ac}+\sigma_{(a,b)}\partial_{a}\epsilon^{c}g_{bc}\,.
\end{align}
Here $\sigma_{(a,b)}=-1$ when both $a,b$ are fermionic indices, and 1 otherwise. This preserves the graded-symmetric nature of $g_{ab}$. One can easily check that under OSp rotations or translations the transformation laws~\eqref{diff-trans} reduce to the expected ones (see e.g. \cite{Kaviraj:2020pwv}, section 3).
Plugging the transformations~\eqref{diff-trans} into~\eqref{var-ac} leads to the following conservation equations:
\begin{align}
0&=\int \text{d}t\,\text{d}^{2|2}y\big(\frac{1}{2} T^{ab}(\partial_{b}\epsilon^{c}g_{ac}+\sigma_{(a,b)}\partial_{a}\epsilon^{c}g_{bc})+J^{a} (\epsilon^{b}\partial_{b}A_{a}+\partial_{a}\epsilon^{b}A_b) \big)\nonumber\\
&=\int \text{d}t\,\text{d}^{2|2}y\big(- \frac{1}{2} (\partial^{b}{T^a}_{b})g_{ac}\epsilon^c+J^{a} \big((\partial_{b}A_{a})-\sigma_{(a,b)} (\partial_{a}A_{b}) \big)\epsilon^{b} + (\partial_0\rho) \epsilon^b A_b \big)
\end{align}
In the second line we used integration by parts and used $\partial^a J_{a}=-\partial_0 \rho$. We can simply ignore the last term in the second line since it is a total derivative in $\partial_0$ if $\epsilon^a$ and $A_a$ do not depend on $t$.\\ \\
Now if we use $\partial_a  A_{b}=- B/2 \, \varepsilon_{ab}$, we simply get 
\begin{align}\label{cons2}
\boxed{\partial^{c}{T^{a}}_cg_{ab} + B J^{a}\varepsilon_{ba}=0\,.}
\end{align}
This is the supersymmetric analogue of the `force balance' condition of the pure model. It should be understood to hold only when we fix the gauge to $A_a=B/2 \, \varepsilon_{ab}y^b$\,, which corresponds to setting $B = B^{'}$. Recalling the discussion in section \ref{sec:specialgauge}, this is a necessary condition to have the full Parisi-Sourlas SUSY emerging in the LLL. 

\subsection{Supercurrents at LLL}\label{sec:supcurr}
Let us take a supercurrent $\mathcal{O}=\rho, J^a, T_{ab}$, defined in subsection \ref{supercurrents}, and consider a general correlation function $\langle \mathcal{O}\cdots\rangle$. These correlation functions must respect a set of Ward identities from the conservation equations~\eqref{cons1} and~\eqref{cons2}.

We saw in~\eqref{susy-X1X2} that in order to take the limit $m\to 0$ one has to introduce the auxiliary fields $X_1, X_2$\,. Above we derived the conservation equations for the supercurrents without these auxiliary fields. However, inside correlation functions, i.e. in the path integral, $\mathcal{O}$ can be written as a variation of $\exp [-S]$ with respect to a source term $J_{\mathcal{O}}=\{A_0, A_a, g_{ab}\}$. If we introduce $X_1, X_2$ in the path integral it becomes clear that the same Ward identities are satisfied if the definitions of the super-currents are modified to 
\begin{align}
\rho &:=\frac{\delta }{\delta A_0}S[g,A,\Phi,X_1,X_2]\,, \\
J^a &:=\frac{\delta }{\delta A_a}S[g,A,\Phi,X_1,X_2]\,. \\
T_{ab} &:=2 \frac{\delta }{\delta g_{ab}}S[g,A,\Phi,X_1,X_2]\,.
\end{align}
We can safely proceed with the computation of the supercurrents in terms of $\Phi, X_1, X_2$. The restriction to the SUSY LLL theory can then be obtained by taking the limit $m\to 0$. Since the SUSY LLL theory can be mapped to the LLL limit of the disordered model~\eqref{eq:org-dis}, the correlators of the supercurrents also get mapped to disordered observables. The SUSY Ward identities then imply a set of new Ward identities for these disordered observables in the LLL limit. We discuss this in the next subsection. \\ \\
The full action~\eqref{susy-X1X2} with the complete dependence on  $A_0, A_a$ and $g_{ab}$ is given by 
\begin{align}\label{susy-X1X2-1}
S[\Phi,X_1,X_2]=\int \td t\td^{d|2}y\, \sqrt{g}\, &\Big[i\Phi^{\dagger}{D}_0 \Phi -  X_1^\dagger(\bar{e}^iD_i\Phi) - (e^i \widetilde{D}_i\Phi^\dagger) X_1 +  X_2^\dagger(\bar{e}^p D_p\Phi)+ (e^p \widetilde{D}_p\Phi^\dagger) X_2\nonumber\\
& + \frac{m}{2} (X_1^\dagger X_1- X_2^\dagger X_2)+ \frac{g^{ip}}{2m}(\widetilde{D}_{(i}\Phi^{\dagger}) D_{p)}\Phi+V(\Phi) \Big]\,,
\end{align}
with $D_0 = \partial_0-iA_0\,,  D_a=\partial_a-iA_a$\,. Varying with respect to $A_0$ and $A_a$ we get $\rho$ and $J^a$ respectively. They are given by
\be\label{susy-j}
\rho = \Phi^{\dagger} \Phi\,, \quad J^{i} = i \bar{e}^i X_{1}^{\dagger} \Phi - i e^i \Phi^{\dagger} X_1\, \quad J^{p} = -i \bar{e}^p X_{2}^{\dagger} \Phi + i e^p \Phi^{\dagger} X_2 \:.
\ee
The derivation of the stress-tensor $T_{ab}$ is more involved, due to the presence of the Zweibeine $e^a, \bar{e}^a$ in the action~\eqref{susy-X1X2-1} which implicitly depend on the metric. In the case where $a,b$ are both bosonic or both fermionic, one can write the general metric in terms of the Zweibeine in the following way:
\begin{align}
g^{ij}&=2(e^i\bar{e}^j+\bar{e}^i e^j)\,, \quad g^{pq} = 2 (\bar{e}^pe^q-e^p \bar{e}^q)\,. \label{eq:metricId}
\end{align}
The off-diagonal terms $g^{ip}$, with one bosonic and one fermionic index, appear explicitly in~\eqref{susy-X1X2-1}. We are ultimately interested in the case where $g^{ip} = 0$, but for the derivation of the stress-tensor it is important to keep this term nonzero for the moment.

Inserting the identity written as $2 e^i \bar{e}_i, 2 e^p \bar{e}_p$ allows us to reinstate the metric in~\eqref{susy-X1X2-1} and determine the components $T_{ij}, T_{pq}$. This is further specified and executed in appendix~\ref{ap:susySTWI}. The result is
\begin{align}\label{susy-T}
        T_{ij} &= -2 \bar{e}_j X^{\dagger}_1 D_i \Phi - 2 e_j (\widetilde{D}_i \Phi^{\dagger}) X_1 - g_{ij} \Big[2 V(\Phi) - \Phi^{\dagger} \frac{\delta V(\Phi)}{\delta \Phi^{\dagger}} -  \frac{\delta V(\Phi)}{\delta \Phi} \Phi  \Big] \:, \\
    T_{pq} &= 2 \bar{e}_q X^{\dagger}_2 D_p \Phi +2 e_q (\widetilde{D}_p \Phi^{\dagger}) X_2 -  g_{pq} \Big[2 V(\Phi) - \Phi^{\dagger} \frac{\delta V(\Phi)}{\delta \Phi^{\dagger}} -  \frac{\delta V(\Phi)}{\delta \Phi} \Phi \Big] \:.
\end{align}
Now consider the case of $T_{ip}$, with one bosonic and one fermionic index. Although simple to evaluate it is singular in the $m\to 0$ limit:
\be\label{crossedT}
T_{ip}=\frac{1}{m}(\widetilde{D}_{(i}\Phi^{\dagger}) D_{p)}\Phi\,.
\ee
Even though the $1/m$ term seems problematic when taking the LLL limit, we will see that is possible to take the $m \to 0$ limit safely using the equations of motion. 
Let us recall~\eqref{eq:lll-sp} - \eqref{eq:lll-theta} where we saw that in the $m\to 0$ limit the auxiliary fields become Lagrange mutipliers, resulting in the SUSY LLL conditions. For nonzero $m$ varying the action~\eqref{susy-X1X2} with respect to $X_1, X_2$ gives us the following equations: 
\begin{align}
      &\text{EOM}(X_1^{\dagger})\text{ : \ \ } \bar{e}^iD_i\Phi =- \frac{m}{2} X_1 \quad \quad \ \text{EOM}(X_1)\text{ : \ \ } e^i \widetilde{D}_i\Phi^\dagger = - \frac{m}{2} X_1^\dagger\label{eom-X1}\\
    & \text{EOM}(X_2^{\dagger})\text{ : \ \ }  \bar{e}^pD_p\Phi = - \frac{m}{2} X_2  \quad \quad \text{EOM}(X_2)\text{ : \ \ } e^p \widetilde{D}_p\Phi^\dagger = - \frac{m}{2} X_2^\dagger
   \,.\label{eom-X2}
\end{align}
One can immediately see that contracting $T_{ip}$ with $\bar{e}^i \bar{e}^p$ or $e^i e^p$ eliminates the $m$ dependence, while contracting it with $e^i \bar{e}^p$ or $\bar{e}^i e^p$ gives zero (up to total derivatives) in the $m \to 0$ limit.

In order to satisfy the Ward identities we have to relate the fields $X_1, X_2$ to $\Phi, \Phi^\dagger$. This is possible using the equations of motion of $\Phi$ and $\Phi^\dagger$. When $m\to 0$ the latter gives
\be\label{eom-phi}
\text{EOM}(\Phi^{\dagger}) \text{ : \ \ } i &D_0 \Phi + e^i D_i X_1 - e^p D_p X_2  + \frac{\delta}{\delta \Phi^{\dagger}}\int \text{d}t\text{d}y V(\Phi)=0\,.
\ee
Varying with respect to $\Phi$ gives the conjugate of the above equation. Notice that if $\Phi$ satisfies the SUSY LLL conditions~\eqref{eq:lll-sp}-\eqref{eq:lll-theta}, then so should $D_0 \Phi$ in~\eqref{eom-phi}. This allows us to to eliminate $D_0 \Phi$ by the action of $\bar{e}^i D_i$ or $\bar{e}^p D_p$ on~\eqref{eom-phi}. This gives
\begin{align}
\bar{e}^iD_i\Big[e^j D_j & X_1 - e^p D_p X_2  + \frac{\delta}{\delta \Phi^{\dagger}}\int \text{d}t\text{d}y V(\Phi)\Big]\nonumber \\&=\bar{e}^pD_p\Big[e^i D_i X_1 - e^q D_q X_2  + \frac{\delta}{\delta \Phi^{\dagger}}\int \text{d}t\text{d}y V(\Phi)\Big]=0\:.\label{eom-X-eom-phi}
\end{align}
We end this subsection by pointing out the relation between supercurrents evaluated here and those in \eqref{sup-current}:
\be
(T_{ij})_{\text{LLL}}=(T_{ij})_{\text{SUSY LLL}}|_{\theta, \thetab \to 0}\,, \ \ (J_i)_{\text{LLL}}= (J_i)_{\text{SUSY LLL}}|_{\theta, \thetab \to 0}\,.
\ee

\subsection{Ward identities in the SUSY LLL theory}
In this subsection we will clarify how the supercurrents found above correspond to certain  disorder-averaged observables. This would imply a new set of Ward identities for observables in the disordered theory that are valid at the LLL. 

The superfield $\Phi$ and its components can be related to the chiral superfield $U$ in~\eqref{hol-supfield} at $m=0$. However, the supercurrents as written in~\eqref{susy-j} and~\eqref{susy-T} depend on the auxiliary superfields $X_1^{\dagger}, X_1$ and $X_2^{\dagger}, X_2$ which at the moment cannot be related to any observable in the disordered model~\eqref{eq:org-dis}. The only field that can  be is $\Phi$. 

We now show that it is possible to relate  $X_1, X_2$ to $\Phi$ using the equations of motion~\eqref{eom-X1}, \eqref{eom-X2}, and~\eqref{eom-phi}. For $m=0$ they can be written more compactly by introducing two-component vector differential operators as follows: 
\begin{align}
&\text{D}_\alpha\Phi=0\,,\label{eom1}\\
&i \partial_0\Phi+ \bar{\text{D}}_{\alpha}\text{X}_\alpha+2\lambda(\Phi^{\dagger}\Phi)\Phi=0\,,\label{eom2}\\
&\ \text{D} :=\left(
\begin{array}{c}
D_{\zb} \\
 - D_{\thetab} \\
\end{array}
\right)\,,\ \bar{\text{D}}:=\left(
\begin{array}{c}
 D_{z} \\
-D_{\theta} \\
\end{array}
\right)\,, \ \text{X}:=\left(
\begin{array}{c}
 X_1\\
X_2 \\
\end{array}
\right)\,.
\end{align}
Here $\alpha=1,2$ denotes the indices of $\text{D},\bar{\text{D}}$ and $\text{X}$. We have set $A_0 = 0$ for simplicity and taken $V(\Phi) = \lambda (\Phi^{\dagger} \Phi)^2$.

Starting from an equation like~\eqref{eom2} one can remove the time-derivative term using the SUSY LLL conditions \eqref{eom1} and solve for the auxiliary fields in terms of the interaction potential. For the solving part we will closely follow a method shown in  \cite{Nguyen:2014uea} for the non-SUSY model with an LLL projector  \cite{doi:10.1142/S0217979297001301, PhysRevB.55.R4895}.  Let us define two projector operators in the following way: 
\begin{align}
    P_z &= \sum_{n=0}^{\infty}\frac{1}{n!(B/2)^n} D_z^n D^n_{\zb}\,,\\
    P_{\theta}&=1+\frac{2}{B'} D_{\theta}D_{\thetab}\,.
\end{align}
The operator $P_z$ projects any function onto the bosonic part of our SUSY LLL condition, i.e. $D_{\zb}P_z \, f(z,\zb)=0$. The $P_{\theta}$ projects onto the fermionic part, i.e. $D_{\thetab}P_\theta \,  g(\theta,\thetab)=0$. It is easy to check the both operators satisfy the projector property:
\be
P_zP_z f(z,\zb)=P_z f(z,\zb)\:, \quad P_\theta P_\theta g(\theta,\thetab)=P_\theta g(\theta,\thetab)\,.
\ee
Now let us define a super-projector $P_{\text{susy}}$ that is simply the product of these two and can be written compactly as follows (setting $B = B'$): 
\begin{align}\label{Psusy}
&P_{\text{susy}}:=P_zP_{\theta}=\sum_{n=0}^{\infty}\frac{1}{n!(B/2)^n} \prod_{i=1}^n\bar{\text{D}}_{\alpha_i}\, \prod_{j=1}^n{\text{D}}_{\alpha_j}
\end{align}
In the above expression there is a summation implied over every repeated $\alpha_i$ index. 
The operator $P_{\text{susy}}$ projects any function $F(z,\zb,\theta,\thetab)$ to the SUSY LLL equations, so that $\text{D}_{\alpha}P_{\text{susy}}F(z,\zb,\theta,\thetab)=0$\,. 
Let us repeat that if $\Phi$ obeys the SUSY LLL conditions, then so does $D_0 \Phi$. Then the action of $\text{D}_{\alpha}$ on~\eqref{eom2} gives 
\be\label{eqstosolve}
\text{D}_{\alpha} (\bar{\text{D}}_{\alpha_1} \text{X}_{\alpha_1}  -  2 \lambda(\Phi^\dagger \Phi)\Phi\big) =0\,.
\ee
Since $\text{D}_{\alpha}(P_{\text{susy}}(\Phi^\dagger \Phi)\Phi)=0$\,, we can substitute $2 \lambda (\Phi^{\dagger} \Phi) \Phi \to (1 - P_{\text{susy}}) 2 \lambda (\Phi^{\dagger} \Phi) \Phi$. Then~\eqref{eqstosolve} becomes
\be
\text{D}_{\alpha} \bar{\text{D}}_{\alpha_1} \text{X}_{\alpha_1} - 2\lambda\sum_{n=1}^{\infty}\frac{1}{n!(B/2)^n}\text{D}_{\alpha}\prod_{i=1}^n\bar{\text{D}}_{\alpha_i}\, \prod_{j=1}^n{\text{D}}_{\alpha_j}((\Phi^\dagger \Phi)\Phi)=0\,.
\ee
This implies the following solution for X: 
\be\label{X-result}
\text{X}_\alpha=2\lambda\sum_{n=1}^{\infty}\frac{1}{n!(B/2)^n} \prod_{i=1}^{n-1}\bar{\text{D}}_{\alpha_i}\, \prod_{j=1}^{n-1}{\text{D}}_{\alpha_j}{\text{D}}_{\alpha}((\Phi^\dagger \Phi)\Phi) \,,
\ee
or written more explicitly:
\be
\text{X}_\alpha = 2\lambda\Big[\frac{{\text{D}}_{\alpha}}{(B/2)} + \frac{\sum_{\alpha_1}\bar{\text{D}}_{\alpha_1}{\text{D}}_{\alpha_1}}{2 (B/2)^2}{\text{D}}_{\alpha}+ \frac{\sum_{\alpha_1,\alpha_2}\bar{\text{D}}_{\alpha_1}\bar{\text{D}}_{\alpha_2}{\text{D}}_{\alpha_1}{\text{D}}_{\alpha_2}}{6 (B/2)^3}{\text{D}}_{\alpha}+\cdots\Big]\, (\Phi^\dagger \Phi)\Phi\,.
\ee
We could add to this solution a function $c_\alpha(z,\zb)$ that is a general solution to the vector differential equation $\text{D}_{\alpha_1} \bar{\text{D}}_{\alpha}c_{\alpha}(z,\zb)=0$. These are usually fixed from boundary conditions on $X_\alpha$. We set $c_\alpha=0$\,, so that when the interaction strength $\lambda=0$ one has $X_\alpha=0$. The motivation for this choice is that without the interaction, the theory restricted to the LLL vanishes since the ground state energy is zero. All local operators such as the supercurrents and the super-stress tensor should vanish accordingly.

The expression~\eqref{X-result} provides a way to relate composite operators involving $X_1, X_2$ (e.g. the supercurrents $J, T$) to be mapped to the disorder problem. Following our results in section \ref{sec:gensusy}, only the components of $\Phi$ can be mapped to the fields $\varphi,\psi$ that describe the physics of disordered electron gas.  In this context recall from section \ref{rep-ferm} that Sp$(2)$-singlet correlation functions of $\varphi,\psi$ in the global OSp theory can be written as disorder averaged correlators. Hence, Sp$(2)$-singlet sectors of correlation functions consisting only of $\Phi$ fields can also be mapped to disorder averaged observables.  

\section{Outlook}\label{sec:outlook}

\paragraph{Summary:}
In this paper, we revisited the emergence of a chiral supersymmetry and  dimensional reduction in the LLL of a $2 + 1$-dimensional free electron gas with a disorder potential, first described in \cite{Wegner83} and \cite{BREZIN198424}. Our main result is the formulation of a magnetic Parisi-Sourlas theory given by \eqref{eq:susy-gen}, which is a nonrelativistic and SUSY-deformed version of a usual PS theory. We have highlighted the important properties of our model by the 
boxed equations: \eqref{result1}, \eqref{result2} and \eqref{result3} - where we connect the magnetic PS to the disorder-averaged theory, \eqref{dimred-gen} - on dimensional reduction of correlation functions, and \eqref{cons1}, \eqref{cons2} - on conservation equations for supercurrents. 

The magnetic PS model is  characterized by two parameters: an external SUSY gauge field $A_a$ and a mass $m$.  We show that in the massless limit it gets constrained by a set of SUSY LLL conditions. This limit corresponds exactly to the chiral SUSY theory of \cite{BREZIN198424} emerging in the usual LLL (massless) limit of the disordered electron gas, but in our formulation this happens in a gauge invariant way. The massless limit is special, since it allows for dimensional reduction of a large sector of observables to a 1d quantum mechanics problem. The SUSY formulation also extends to higher Landau levels, since there is a special choice of the SUSY gauge field that maps the magnetic PS model to the disorder one for any mass. Finally we show that there exists a set of supercurrents with super-conservation equations, that amounts to new Ward identities for disorder averaged observables. \\ \\
There are a number of interesting and immediate directions that one could persue within the framework of this paper. We discuss them below. \\

\paragraph{Ward identities in terms of disorder averaged observables:} The SUSY Ward identities established in section \ref{sec:Wardid} should manifest in terms of averaged correlators in the disorder Landau level problem. 
Consider the one-point functions of $T^{ab}$ and $J^{a}$. Both these supercurrents are composite operators of $\Phi$ as well as $X_1, X_2$ 
using the results~\eqref{susy-j} and~\eqref{susy-T}. In order to map them to disorder averaged observables, which should only depend on $\Phi$, we must use~\eqref{X-result}. The part of the one-point function of $T_{ij}$ coming from for example $D_i\Phi^\dagger X_1$  would look like
\be
\langle (D_i\Phi^\dagger X_1)(y)\rangle =  \lim_{y'\to y} 2\lambda\sum_{n=1}^{\infty}\frac{(-\frac 4B)^n}{n!}\, D_i{(y')} \, \prod_{i=1}^{n-1}\bar{\text{D}}_{\alpha_i}(y)\, \prod_{j=1}^{n-1}{\text{D}}_{\alpha_j}(y){\text{D}}_1(y)  \, \langle \Phi^\dagger(y')(\Phi^\dagger \Phi)\Phi(y)\rangle\,.
\ee
It is now straightforward to see that the one-point function of any supercurrent $\mathcal{V}=T,J$ can be written in terms of a four-point correlation function:
\be\label{TJ-dis}
\langle \mathcal{V}(y)\rangle= \lim_{y_i\to y} \, \mathfrak{D}(y_1,y_2,y_3,y_4)\, \langle \Phi^\dagger(y_1)\Phi^\dagger(y_2)\Phi(y_3)\Phi(y_4)\rangle \,,
\ee
where $\mathfrak{D}$ takes care of the various composites in~\eqref{susy-j} and~\eqref{susy-T}. The four-point function $\langle \Phi^\dagger\Phi^\dagger\Phi\Phi\rangle$ packs together various correlators of $\varphi$ and $\psi$. It is possible to take combinations of~\eqref{TJ-dis} so that the correlators in the r.h.s. is an Sp$(2)$ singlet. As long as we focus on that sector the $\vf,\psi$ correlators can be related to disorder averaged ones (see section \ref{rep-ferm} and App. \ref{app:map}). If such observables can be measured in a simulation or experiment, a map like~\eqref{TJ-dis} will provide a way to explicitly check the SUSY Ward identities.

\paragraph{Linear response theory:} An important characterization of a system of charged particles is how it reacts to an external electric or magnetic field. Typically the charge or current densities have a linear dependence on the external fields with the constants of proportionality termed as conductivities or collectively a conductivity tensor. E.g. in the ordinary electron gas problem without disorder potential the one-point function of the current is assumed to depend on the electric field as $\langle J_i(x)\rangle=-\int \td^3x'\,\sigma_{ij}(x-x') \, E_j(x') $. Linearizing with respect to $A_i$ one gets:
    \be\label{jj-sigma}
    \langle J_i(x)J_j(0)\rangle=\partial_0 \sigma_{ij}(x)\,.
    \ee
Here we have used $E_i=\partial_0 A_i$ and $\delta\langle J_i(x)\rangle/\delta A_j(0)=\langle J_i(x)J_j(0)\rangle$.

In the presence of disorder the problem gets mapped to a SUSY theory where a superfield interacts with an external super-gauge field. It is then natural to ask what is the generalization of the linear response. By extending \eqref{jj-sigma} one can define a generalized SUSY conductivity tensor 
    \be\label{susy-sigma}
    \langle J_a(x)J_b(0)\rangle=\partial_0 \sigma_{ab}(x)\,.
    \ee
Let us comment on why relations like \eqref{susy-sigma} could be interesting.\footnote{Note that the fermionic components of $\sigma_{ab}$ determine the response to the fermionic components of the gauge field $A_a$ whose strength, as discussed in the special case II from subsection \ref{higherLL}, determines how far we are from the LLL. Although in
our analysis these components cannot have complete freedom since we have to fix gauge as in \eqref{gauge-choice} to connect to the disordered theory.}
Firstly, they define new transport coefficients that characterize the entire disordered system. Just like $J_a$, correlators of $T_{ab}$ can define SUSY analgoues of elastic modulus and viscosity \cite{Gubser:2008sz,Bradlyn:2012ea}. They should satisfy special relations between themselves, just like their non-SUSY analogues in usual hydrodynamic setup \cite{Geracie:2014nka}. Whether disorder averaged observables are consistent with such conditions can be tested both theoretically and experimentally. 
    
Secondly, computing some of these quantities in the LLL limit could be simpler due to the dimensional reduction property of Parisi-Sourlas theories (see section \ref{dimredsec}). Of course, computing correlation functions of the currents $J_a$ or $T_{ab}$ may involve all the fields $\Phi, X_1, X_2$ and an exact and complete dimensional reduction cannot be expected to hold, but the simple nature of the action \eqref{eq:susy-gen-met} indicates some simplification could be expected (see next point). 

\paragraph{Higher Landau levels:} One of the highlights of this work is that the SUSY proposal \eqref{eq:susy-gen} can be mapped to the disordered Landau quantization problem for nonzero mass $m$. Recall that $m>0$ captures the effect of higher Landau levels. Although dimensional reduction holds only in the $m\to 0$ LLL limit, and for a certain class of observables (see \eqref{dimred-gen}), SUSY is expected to impose some interesting constraints even for nonzero $m$. 

To better establish this point consider the SUSY action written with the $X_1, X_2$ fields. As shown in \eqref{spn-PS} we can separate the purely OSp$(2|2)$ part from the rest of the theory as follows: 
\be
S[\Phi]= S_{\text{PS}}[\Phi] + \big(- X_1^\dagger(\bar{e}^i D_i\Phi) +  X_2^\dagger(\bar{e}^p D_p\Phi)\big)+\text{c.c.}  + \frac{m}{2} (X_1^\dagger X_1- X_2^\dagger X_2)\,.
\ee
One may set up a Feynman diagram computation for a correlator, say with only $\Phi$ and $\Phi^\dagger$ as external fields for simplicity. For  $m=0$ the result is dimensionally reduced as argued in section~\ref{sec:gensusy}. Now one could additionally obtain the leading correction for small $m$. 
It would be interesting to see if the following features appear in these corrections:
\begin{enumerate}
    \item Dimensional reduction, although no longer strictly valid,  should have some of its signatures still visible. To see this one could work with the symmetric gauge of $A_a$ which will involve $\varepsilon_{ij}$ and $\varepsilon_{pq}$ tensors in the internal legs/vertices of Feynman diagrams. Contracted with $g^{ij}$ and $g^{pq}$ respectively they would vanish leading to various simplifications in the  small $m$ corrections.      

    \item For $B'\gg B$ the components of the superfield $\Phi$ are identified with $\varphi$ and $\psi$ in \eqref{eq:ac-with-B}. Therefore SUSY will imply a relation between correlators of $\varphi$'s and $\psi$'s. The correlators can be mapped to distinct disorder averaged observables (see discussion around \eqref{eq:2point-triv}), which therefore must satisfy a nontrivial relation.
\end{enumerate}

\subsection{Future directions}
The physics of non-interacting electrons confined to Landau levels in a strong magnetic field, although a simple setup, is connected to important topics in statistical physics. It provides a way to understand the Quantum Hall Effect (QHE) and related problems.
In high energy physics, such systems are often studied in connection to supersymmetric CFTs, hydrodynamics, and more.
Our results open up a number of future directions in all these avenues. Below we list a few of them.
\begin{enumerate}
    \item \emph{Hall effect and localization}. 
    The theory of localization \cite{PhysRevLett.42.673} says that a system of non-interacting electrons in the presence of disorder has localized states  which prevents a metallic phase (conduction). This happens even in weak disorder for $d\leq 2$. But the situation changes when there is a magnetic field - in the continuum of the energy spectrum there are discrete critical energy levels corresponding to extended (non-localized) states that result in nonzero conductivity: the Integer Quantum Hall Effect (IQHE) \cite{tong2016lecturesquantumhalleffect}. The so-called localization length diverges as a power law as the energy approaches a critical value, but it is not properly understood if this divergence is universal \cite{local2,local3}. Turning on inter-electron interactions further complicates the theory \cite{local1}, and gives rise to the Fractional Quantum Hall Effect (FQHE). The metal-insulator transition at zero magnetic field is also an interesting mechanism to understand \cite{localmetins}. Mapping the disorder to a SUSY theory may grossly simplify computations of observables in many of the above directions. 

    \item \emph{Laughlin states and conformal blocks}.
    Introducing interactions between electrons allows one to study the FQHE, but at the same time complicates the current setup. The wavefunctions of the electrons in the Landau levels, will be more complicated as well.
    The Ansatz for these wavefunctions is given by Laughlin wavefunctions \cite{Laughlin:1983fy}. In \cite{Moore:1991ks}, it is shown that there exist a connection between Laughlin wavefunctions and conformal blocks for rational CFTs in $1 + 1$ dimensions. Note that this connection is not a result of criticality of the system, but relies on certain braiding properties being present in both the FQHE system and $1+1$-dimensional rational CFTs.  
    
    Incorporating interactions between electrons will require a substantial modification of our current field-theoretic description. Nevertheless, if such a modification can be made, it might be possible to show a relation with the conformal blocks, related to Laughlin wavefunctions, used to describe the ground state. 
    
    \item \emph{Time-dependent diffeomorphism invariance}.
    In nonrelativistic systems sometimes a larger symmetry than usual spatial diffeomorphisms can be identified. This symmetry corresponds to a time-dependent reparameterization of coordinates, and leads to new conserved currents. Furthermore, in the context of quantum Hall physics, this leads to a Weyl invariance in the LLL \cite{Geracie:2014nka}. It would be interesting to show that  SUSY extensions of the enhanced symmetries exist for our model. In particular it would be nice to formulate it as a generalization of the Newton-Cartan geometry \cite{ASENS_1923_3_40__325_0,ASENS_1924_3_41__1_0}, similar to \cite{Son:2013rqa,SON2006197}. 

    \item \emph{Chern-Simons theories}.
    Instead of looking at the physics of Landau levels, the QHE can also be explained with the help of topological quantum field theories in $2 + 1$ dimensions, called Chern-Simons (CS) theories. The level of the CS term corresponds to the filling factor of the Landau levels, and hence gives the conductivity for both the IQHE and FQHE \cite{FROHLICH1991517,Lopez91}. The CS theory gives an elegant topological explanation of the QHE without relying on the microscopic physics. It would be interesting to see if we can write a dual description of our system in terms of a SUSY CS theory. Work on SUSY CS theories related to quantum Hall physics has already appeared in \cite{Leblanc:1992wu,Tong:2015xaa}

    \item \emph{Kinematics of dimensional reduction}. 
    In a Parisi-Sourlas SUSY CFT$_d$ correlation functions dimensionally reduce to a CFT$_{d-2}$. Recently it has been shown that under special conditions, the equivalence can also be reversed, i.e. a $(d-2) \to d$ uplift of correlation functions \cite{Trevisani:2024djr}. In CFTs such relations imply special identities between conformal blocks or even Witten diagrams of different dimensions \cite{Kaviraj:2019tbg,Zhou:2020ptb,Hoback:2020syd}. Dimensional reduction under disorder in the LLL implies the existence of interesting inter-dimensional identities, especially for the CFTs studied in connection to the quantum Hall effect. In the light of our new superfield picture, such conformal identities would be a fascinating new direction to explore. 
\end{enumerate}

\section*{Acknowledgments}
We thank Adhip Agarwala, Ant\'onio Antunes, \'Edouard Br\'ezin, Diptarka Das, Sumit Das, Arijit Kundu, Gautam Mandal, and Alessio Miscioscia for useful discussions. We also thank \'Edouard Br\'ezin, Gautam Mandal, Miguel Paulos, Emilio Trevisani, and Slava Rychkov for their comments on the draft. PvV is funded by the European Union (ERC, FUNBOOTS project, project number 101043588, held by M. Paulos). Views and opinions expressed are however those of the authors only and do not necessarily reflect those of the European Union or the European Research Council Executive Agency. Neither the European Union nor the granting authority can be held responsible for them.

\appendix

\section{Disordered Fermionic gas}\label{app:fermion-dis}

In the main text we started with a disordered model that is a theory of bosons coupled to a magnetic field and a disorder interaction. The purpose of this appendix is to show that starting with fermions instead, leads us to the same theory as bosons after averaging out the disorder. The eventual emergence of the chiral SUSY at LLL limit and the map to a magnetic PS model is remains unchanged. \\ \\
Let us start with the following disordered model:
\be\label{eq:org-dis-ferm}
S[\fsi,h]=\int \text{d}^2x \text{d}t \Big(i {\fsi}^\dagger D_0  \fsi -\frac{1}{2m}|D_i \fsi|^2 +\frac{B}{2m}\fsi^\dagger\fsi+ h(t,x)\,\fsi^\dagger\fsi\Big)
\ee
Here $\fsi$ is a complex spinless fermionic field. We assume that the disorder $h(t,x)$ is quenched and is drawn from a distribution with the same properties as \eqref{eq:avgdef}. For simplicity we take 
\be
\overline{h(t,x)h(t',x')}=\lambda\, \delta(x-x')\delta(t-t')\,.\label{eq:avgdefapp}
\ee
We remind the reader that more general time dependence is allowed for $h(t,x)$ at the cost of a nonlocal interaction after averaging.

To carry out disorder averaging we follow the steps of sec. \ref{repac} i.e. we write the path integral representation of a generic quenched-averaged observable $\langle A(\fsi)\rangle$ and insert the partition function $Z[h]^{n-1}$ in both numerator and denominator. This gives
\begin{align}\label{ferm-repl}
\overline{\langle A(\fsi)\rangle}
&= \int\mathcal{D}h \,  \mathcal{P}(h) \,  \frac{\int \mathcal{D}\fsi_1\cdots\mathcal{D}\fsi_n \, A(\fsi_1)\, e^{-\sum_{\a=1}^n S[\fsi_\a,h]}}{\big(\int \mathcal{D}\fsi \, e^{ -S[\fsi,h]}\big)^n}\,.
\end{align}
Next we assume this has a smooth limit as $n\to 0$ so that the denominator of \eqref{ferm-repl} becomes 1. This allows us to carry out the $h$ average which gives 
\be
S_{\text{rep}}[\fsi_\a]=& \int \text{d}^2x \text{d}t\left[i \sum_\a \big({\fsi^{\dagger}_\a}D_0\fsi_\a - \frac{1}{2m} |D_i\fsi_\a|^2+\frac{B}{2m}\fsi^{\dagger}_\alpha\fsi_\alpha\big)+ \lambda\Big(\sum_\a\fsi_\a^\dagger \fsi_\a\big)^2\right] \label{eq:rep-action-app}
\ee
The interaction term could be written as a quadratic term with an auxiliary field using Hubbard-Stratonovich transformation as follows:
\be
\int \text{d}^2x \text{d}t \Big[\lambda (\fsi^\dagger_\alpha\fsi_\alpha)^2\Big] \to \int \text{d}^2x \text{d}t \Big[i\xi\fsi^\dagger_\alpha\fsi_\alpha+\frac{\xi^2}{4\lambda}\Big]
\ee
Now we may isolate the replica fields $\fsi_1$ from the rest of the replica fields in the resulting replica action $S_\text{rep}[\fsi_\alpha,\xi]$.  Then the Gaussian path integral of $n-1$ fermionic replica fields can be written as
\be\label{PIident-app}
\int\, \prod_{\a=2}^n\mathcal{D}\fsi_\a\exp\big[- \sum_{\a=2}^{n}\fsi^\dagger_\a \,\text{M}\, \fsi_\a\big] \ =\left(\frac{\text{det M}}{\pi}\right)^{n-1} \ \  \stackrel{n\to 0}{=}\int\mathcal{D}\varphi\, \exp\big[- i \varphi^\dagger \text{M} \varphi\big]\,,
\ee
where $\varphi$ is a complex scalar boson. Now if we integrate out $\xi(t,x)$ and rename $\fsi_1\to \psi$ then the resulting action looks like
\be\label{eq:bos-fer-app}
S[\varphi,\psi]= \int \text{d}^2x \text{d}t\left[i {\varphi^\dagger}D_0\varphi + i \psid D_0\psi - \frac{1}{2m}( |D_i\varphi|^2+|D_i\psi|^2)+ \frac{B}{2m}(\varphi^\dagger\varphi+\psid\psi)+\lambda(\varphi^\dagger\varphi+\psid\psi) ^2\right]\,.
\ee
Therefore we get back the global OSp theory from \eqref{eq:bos-fer} that describes the bosonic electron gas after disorder averaging. The rest of the analysis of \ref{sec:model} i.e. the emergence of chiral SUSY in the $m\to 0$ limit follows identically. Also our main proposal from sec. \ref{sec:proposal} i.e. the magnetic PS model $S[\Phi]$ reduces to \eqref{eq:bos-fer-app} when we choose the special gauge of \eqref{gauge-choice}. 

In order to access $\psi$ observables from the magnetic PS model one has to focus on the correlators of $\Phi_{\theta}$, since $\Phi_\theta\to \psi$ in the fermionic SUSY LLL limit \eqref{thetalll-sol}. This makes it clear how the SUSY formalism of our paper works for fermionic model in the same way as the bosonic model. 

\section{Mapping between SUSY and disorder averaged observables}\label{app:map}
In this appendix we will clarify the relations between various theories defined in section \ref{sec:model}. For restricted sectors of correlation functions we will show the equivalence between the replica action \eqref{eq:rep-action} and the global supersymmetric theory \eqref{eq:bos-fer}. This will also let us identify the relations between correlation functions of global SUSY fields $\varphi,\psi$ and disorder averaged observables.  

Take the following two theories: one involving $n$ complex scalar fields $\phid_\alpha, \phi_\alpha$ where $\alpha=1,\cdots, n$ and it has a global  O$(n)\times$U$(1)^{n}$ symmetry. The other is for a pair of bosons and fermions, $\varphi^\dagger,\varphi$  and $\psid,\psi$, again with a quadratic action having a global OSp$(2|2)$ SUSY. Let us consider them to have to following nonlocal quadratic actions:
\begin{align}
&S[\phid_\alpha,\phi_\alpha]:=\int\int\td^2x_1\td^2x_2 \, \phid_\alpha(x_1)M(x_1,x_2)\phi_\alpha(x_2)\label{quadrep}\\
&S[\varphi^\dagger,\varphi,\psid,\psi]:=\int\int\td^2x_1\td^2x_2 \, 
\Big(\varphi^\dagger(x_1)M(x_1,x_2)\varphi(x_2)+\psid(x_1)M(x_1,x_2)\psi(x_2)\Big) \label{quadglob}\,.
\end{align}
We suppressed the $t$-integral for simplicity. 
For local theories $M(x,y)$ will be linear combinations of delta function and its derivatives. In this way the replica theory defined in \eqref{eq:rep-action} and the global SUSY theory of \eqref{eq:bos-fer} can be made to fall in these categories respectively. 

In the $S[\phid_\alpha,\phi_\alpha]$ theory let us group the $n$ fields as $\phi_1$ and $\{\phi_2,\cdots ,\phi_n\}$. Then it is easy to see that any correlation function $\langle F(\phid_1,\phi_1)\rangle$ computed in $\eqref{quadrep}$ is identical to the same correlator with fields replaces as $\langle F(\varphi^\dagger,\varphi)\rangle$ computed in \eqref{quadglob}.

The equivalence in the $\phi_\alpha$, $\alpha=2,\cdots, n$ and $\psi$ sectors are not so straightforward. We will now show that an equivalence does exist for a class of correlators in each theory.
Consider the following path integrals for the two theories restricted to the above fields only
\begin{align}
   Z_{\phi}[A(x,y)]= &\int \prod_{\alpha=2}^n\mathcal{D}\phi_\alpha\, e^{-\big[S[\phid_{\alpha},\phi_\alpha]+\int_x\int_y \,\phid_\alpha(x) A(x,y)\phi_\alpha(y)\big]}=\mathcal{N}_{\phi}(\text{det}(M+A))^{-\frac{2n-2}{2}}\,.\\
   Z_{\psi}[A(x,y)]= &\int \prod_{i=2}^n\mathcal{D}\psi\, e^{-\big[S[\psid,\psi]+\int_x\int_y \,\psid(x) A(x,y)\psi(y)\big]}=\mathcal{N}_{\psi}(\text{det}(M+A))\,.
\end{align}
Here we have taken a source $A(x,y)=A(y,x)$ for bilocal operators. In $n\to 0$ limit, taking derivatives of $\log Z_\phi$ and $\log Z_\psi$ w.r.t. $A(x,y)$ and accounting for the $x$-$y$ symmetry one immediately obtains the equality between correlation functions of bilocal operators 
\begin{align}
&\sum_{\alpha=2}^n\langle (\phid_\alpha(x)\phi_\alpha(y)+\phid_\alpha(y)\phi_\alpha(x))\rangle =  \langle (\psid(x)\psi(y)+\psid(y)\psi(x))\rangle\\
&\sum_{\alpha,\beta=2}^n\langle (\phid_\alpha(x_1)\phi_\alpha(x_2)+\phid_\alpha(x_2)\phi_\alpha(x_1))(\phid_\beta(x_3)\phi_\beta(x_4)+\phid_\beta(x_4)\phi_\beta(x_3))\rangle \nonumber\\ & \hspace{1cm}=\langle (\psid(x_1)\psi(x_2)+\psid(x_2)\psi(x_1))(\psid(x_3)\psi(x_4)+\psid(x_4)\psi(x_3))\rangle\,,
\end{align}
and so on. These relations can be more compactly written as
\be\label{ferm-rep-map}
\left\langle f_N\big(\text{Re$\sum_{\alpha=2}^n$}(\phid_\alpha(x)\phi_\alpha(y))\big)\right\rangle =  \left\langle f_N\big(\text{Re}(\psid(x)\psi(y))\big)\right\rangle\,.
\ee
Here $f_N(\mathcal{O}(x,y)):= \prod_{i=1}^{N}\mathcal{O}(x_i,y_i)$\,, so that $N=1$ corresponds to a 2-point function, $N=2$ to a 4-point function, etc. We may use the identity \eqref{ferm-rep-map} to obtain correlation functions of composite local operators. E.g. for the $N=2$ case above, dressing with derivatives and taking appropriate limits  give us
\begin{align}
\langle(\phid_\alpha\partial^2\phi_\alpha+(\partial^2\phid_\alpha)\phi_\alpha)(x)(\phid_\alpha\phi_\alpha)(y)\rangle&=\langle(\psid\partial^2\psi+(\partial^2\psid)\psi)(x)(\psid\psi)(y)\rangle\,,\\
\langle (\partial_i\phid_\alpha\partial_j\phi_\alpha)(x)(\partial_k\phid_\beta\partial_l\phi_\beta)(y) \rangle&=\langle (\partial_i\psid\partial_j\psi)(x)(\partial_k\psid\partial_l\psi)(y) \rangle\,,
\end{align}
etc. So starting from a correlation function of $2N$ fields in the global SUSY theory that is a singlet under the Sp$(2)$ subgroup we can identify it with a replica correlation function of $2N$ replica fields that is a singlet under O$(n-1)\times$U$(1)^{n-1}$. The map also lets us go the other way. 

Let us demonstrate the relation of these correlation functions to  disorder averaged observables with an example: consider the 4-point function $\langle\psid(x_1)\psi(x_2)\psid(x_3)\psi(x_4)$. Using \eqref{ferm-rep-map} we can write it in terms of
\begin{align}
&F(x_1,x_2,x_3,x_4):=\sum_{\alpha,\beta=2}^n\langle\phid_\alpha(x_1)\phi_\alpha(x_2)\phid_\beta(x_3)\phi_\beta(x_4) \rangle\nonumber\\ &= \sum_{\alpha=2}\langle\phid_\alpha(x_1)\phi_\alpha(x_2)\phid_\alpha(x_3)\phi_\alpha(x_4) \rangle
 +  \sum_{i\neq j}^n\langle\phid_\alpha(x_1)\phi_\alpha(x_2)\phid_\beta(x_3)\phi_\beta(x_4) \rangle\,.
\end{align}
Using symmetry in exchange of replicas we can write the last line as
\be
(n-1)\langle\phid_2(x_1)\phi_2(x_2)\phid_2(x_3)\phi_2(x_4) \rangle+ (n-1)(n-2)\langle\phid_2(x_1)\phi_2(x_2)\phid_3(x_3)\phi_3(x_4) \rangle\,.
\ee
Now using relations \eqref{eq:dis-corr-repl} and \eqref{avg-repl-2} from the main text that map disorder-average correlators to the replica theory we can write 
\be
F(x_1,x_2,x_3,x_4)=-\overline{\langle\phid(x_1)\phi(x_2)\phid(x_3)\phi(x_4) \rangle}+ 2\overline{\langle\phid_2(x_1)\phi_2(x_2)\rangle\langle\phid_3(x_3)\phi_3(x_4) \rangle}\,.
\ee
This is how a large class of observables built from $\psid,\psi$ can be mapped to the disorder theory.

When working with operators in the Parisi-Sourlas theory one often encounters operators that are not singlets under Sp$(2)$, e.g. $\Phid\partial_{\theta}\Phi(\thetab=\theta=0)\propto \varphi^\dagger\psi$ if we use the fermionic LLL condition \eqref{eq:lll-theta}. To be meaningful in the disordered theory one can combine them with other operators to form a global Sp$(2)$ singlet as above. E.g. consider the correlation function of the above operator and two other SUSY fields
\be
\langle(\Phid\partial_{\theta}\Phi)(x)\,(\partial_{\thetab}\Phid)(y)\,\Phi(z)+\partial_\theta\Phi(y)(\Phid\partial_{\thetab}\Phid)(x)\Phi(z)\rangle\,.
\ee
It is proportional to $\langle (\varphi^\dagger\psi)(x)\psid(y)\varphi(z)+\psi(y)(\varphi\psi)^\dagger(x)\varphi(z) \rangle$ an Sp$(2)$ singlet correlator in which \eqref{ferm-rep-map} is applicable.

\section{Conventions}
In this appendix we summarize our conventions. 
\subsection{The metric and epsilon tensor}
The flat $OSP$ metric is given by:
\be
g^{(2|2) ab}:=\left(\begin{array}{cc}
 \eta & \ 0 \\
 0 & \ i\,\varepsilon  \end{array}\right)\,, \ \eta^{ij}:= \left(\begin{array}{cc}
 1 & \ 0 \\
 0 & \ 1 \\\end{array}\right)\,, \ \varepsilon^{pq}:=\left(\begin{array}{cc}
 0 & \ 1 \\
 -1 & \ 0 \\\end{array}\right)\,.
\ee
Note that in principle, there could also be off-diagonal elements $g^{ip}$. We will consider them to be zero, but they will play a role in e.g. the derivation of the stress-tensor.
We will also need an epsilon tensor
\be\label{susy-varep-app}
\varepsilon^{(2|2) ab} :=\left(\begin{array}{cc}
\varepsilon & \ 0 \\
 0 & \ i \, \eta  \end{array}\right)\,.
\ee
Again, off-diagonal elements are set to zero.
Note that this tensor is not an invariant tensor of the $OSP$ symmetry group.
These definitions allow us to rewrite~\eqref{eq:action-new} as:
\begin{align}
S[\Phi]&=\int \td^2x \td^2\theta\, \Big[i \Phi^{\dagger}D_0 \Phi- \frac{1}{2m}({(g^{(2|2)}})^{ab}+i {(\varepsilon^{(2|2)}})^{ab})(\widetilde{D}_a \Phi^\dagger)(D_b \Phi)+\lambda(\Phi^\dagger \Phi)^2 \Big]\nonumber\\
&=\int \td^2x \td^2\theta\, \Big[i \Phi^{\dagger}D_0 \Phi- \frac{1}{2m}(\eta^{ij}+ i\varepsilon^{ij})(\widetilde{D}_i \Phi^{\dagger}) D_j \Phi + \frac{1}{2m}( \eta^{pq}-i\varepsilon^{pq})(\widetilde{D}_p \Phi^{\dagger}) D_q \Phi+\lambda(\Phi^\dagger \Phi)^2 \Big]\,, \label{eq:actionRewrite}
\end{align}
Here, $\widetilde{D}_a$ is defined such that the theory is gauge-invariant:
\be
\widetilde{D}_a = \partial_a + i A_a\:. 
\ee
The (flat) $OSP$ metric and the epsilon tensor can be rewritten in terms of delta functions, which can be combined in complex Zweibeine:
\begin{align}
    &e_i = \frac{1}{2} (\delta_{i}^1 - i \delta_{i}^2)\:, \quad \bar{e}_i = \frac{1}{2} (\delta_{i}^1 + i \delta_{i}^2)\:, \quad e^i = \frac{1}{2} (\delta^{1 i} - i \delta^{2 i})\:, \quad \bar{e}^i = \frac{1}{2} (\delta^{1 i} + i \delta^{2 i})\:, \nonumber \\
    &e_p = - \frac{1}{2} (\delta_{p}^3 + i \delta_{p}^4)\:, \quad \bar{e}_p = \frac{1}{2} (\delta_{p}^3 - i \delta_{p}^4)\:, \quad e^p = \frac{1}{2} (\delta^{3p} + i \delta^{4p})\:, \quad \bar{e}^p =  \frac{1}{2} (\delta^{3p} - i \delta^{4p})\:, \nonumber \\
\end{align}
Note that there is an extra minus sign for $e_p, \bar{e}_p$ that does not appear for $e_i, \bar{e}_i$. The $\delta$'s are Kronecker deltas. 
Using these Zweibeine, we can derive identities such as:
\be
(\eta^{ij}+ i \varepsilon^{ij})(D_i \Phi)^{\dagger} D_j \Phi-(\eta^{pq}- i \varepsilon^{pq})(\widetilde{D}_p \Phi)^{\dagger} D_q \Phi \ = \  4 (\bar{e}^iD_i \Phi)^{\dagger} (\bar{e}^jD_j \Phi) - 4 (e^p \widetilde{D}_p \Phi^{\dagger}) (\bar{e}^q D_q \Phi)\,.
\ee

\subsection{Complex coordinates}
We take the following conventions for complex coordinates $z, \bar{z}, \theta, \bar{\theta}$, where $\theta, \bar{\theta}$ are Grassmann-odd variables (but have zero spin):
\begin{align}
    z = x_1 + i x_2\:, \quad \bar{z} = x_1 - i x_2\:, \quad \theta = \theta_1 - i \theta_2\:, \quad \bar{\theta} = \theta_1 + i \theta_2\:. 
\end{align}
The covariant derivatives are written in terms of these coordinates as:
\begin{align}
    &D_z:=e^iD_i=\partial_{z} - \frac{B}{4} \zb\,, \quad D_{\zb} :=\bar{e}^i D_i=\partial_{\zb} + \frac{B}{4} z \\
    &D_{\thetab}:= \bar{e}^p D_p=\partial_{\thetab} -  \frac{B'}{4}\theta \,, \quad \widetilde{D}_{\theta}:= e^p \widetilde{D}_p = \partial_{\theta} + \frac{B'}{4} \thetab\:.
\end{align}
This implies the following conventions for the gauge field $A_a$:
\begin{align}
    A_1 &= - \frac{B}{2} x_2\:, \quad A_2 = \frac{B}{2} x_1\:, \nonumber \\
    A_3 &= -i \frac{B'}{2} \theta_1\:, \quad A_4 = -i \frac{B'}{2} \theta_2\:. 
\end{align}
Note that $A_i$ is real, while $A_p$ is complex. 
Furthermore, the magnetic field $B + B'$ is defined in terms of $A_a$ as: 
\be
    \widetilde{B} = B + B' = \varepsilon^{(2|2) ab} \partial_a A_b\:.
\ee
The superfield $\Phi$ and its complex conjugate can be expanded in $\theta, \thetab$:
\begin{align}
    &\Phi(x,\theta)=\Phi_0(x)+\theta \Phi_{\theta}(x)+\thetab \Phi_{\thetab}(x)+ \theta\thetab \Phi_{\theta\thetab}(x)\,, \label{superfield} \\
    &\Phi^{\dagger}(x,\theta)=\Phi^{\dagger}_0(x)+ \Phi^{\dagger}_{\theta}(x) \thetab + \Phi^{\dagger}_{\thetab}(x) \theta+ \theta\thetab \Phi^{\dagger}_{\theta\thetab}(x)\,.
\end{align}
The LLL conditions (and their complex counterparts) in eq.~\eqref{eq:lll-theta} lead to constraints on these expansions:
\begin{align}
    &D_{\thetab} \Phi = -\frac{B}{4} \theta \Phi^0 + \Phi_{\thetab} - \theta \Phi_{\theta \thetab} = 0 \to \Phi_{\theta \thetab} = - \frac{B}{4} \Phi_0\:, \quad \Phi_{\thetab} = 0\:, \nonumber \\
    &\widetilde{D}_{\theta} \Phi^{\dagger} = \frac{B'}{4} \thetab \Phi^{\dagger}_0 - \Phi^{\dagger}_{\thetab} + \thetab \Phi^{\dagger}_{\theta \thetab} = 0 \to \Phi^{\dagger}_{\theta \thetab} = - \frac{B}{4} \Phi^{\dagger}_0\:, \quad \Phi^{\dagger}_{\thetab} = 0\:.
\end{align}
Imposing these relations means that we can identify $\Phi_{\theta}$ with the $\psi$'s of section~\ref{rep-ferm}, and $\Phi_0$ with $\varphi$. 

\subsection{The super-Poincar\'e algebra}
\label{app:spa}

A pure Parisi-Sourlas theory has super-Poincar\'e symmetry and is invariant under supertranslations and -rotations which preserve the distance $y^2 = g_{ab}^{(2|2)} y^a y^b = x_{1}^2 + x_{2}^2 + 2 i \theta_1 \theta_2$, where $\theta_1,\theta_2$ are Grassmann-odd coordinates. The translations and rotations are given by
\be
y^a \to y^{a'} = y^a + c^a\:, \quad y^a \to y^{a'} = R^{a}_b y^b\:,
\ee
where $c^a$ is a constant and $R^{a}_b$ a rotation matrix. The generators of translations are given by $P_a = \{P_i, P_{\theta_1}, P_{\theta_2}\}$ and can be expressed as usual in terms of derivatives with respect to coordinates in superspace: $P_a = \partial / \partial y^a$. The generators of superrotations are written as $M^{ab}$, where $a,b = i,\theta_1, \theta_2$. Note that $M^{ab}$ is bosonic if both $a,b$ are bosonic or fermionic, and fermionic if $a,b$ are graded differently. The generators $P_a$ and $M^{ab}$ satisfy graded commutation relations, which form the super-Poincar\'e algebra
\begin{align}
    &[P_a, P_b\} = 0\:, \quad [M^{ab}, P^c\} =  g^{cb} P^a - (-1)^{[a][b]} g^{ca} P^b\:, \nonumber\\
    &[M^{ab}, M^{cd}\} = g^{cb} M^{ad} - (-1)^{[a][b]} g^{ca} M^{bd} - (-1)^{[c][d]} g^{db} M^{ac} + (-1)^{[a][b] + [c][d]} g^{da} M^{bc}\:,
\end{align}
where $[a] = 0,1$ for bosonic and fermionic indices respectively.
The notation $[\cdots \}$ indicates graded commutators, and are defined as
\be
[ A,B \} = \begin{cases} [A,B] & \text{ if $A$ or $B$ is Grassmann even,} \\
 \{A,B \} & \text{ if $A$ and $B$ are Grassmann odd.}
 \end{cases}
\ee
See also section 3 of \cite{Kaviraj:2019tbg} for a more extensive review.

\section{The SUSY stress-tensor}
\label{ap:susySTWI}
In this appendix we show how to obtain the currents and stress-tensor from the action of the supersymmetric theory. 
Starting from the action~\eqref{susy-X1X2-1}, we can find the currents and stress-tensor by varying the action with respect to $A_a$ and $g^{(2|2) ab}$ respectively. 
The derivation of the currents is straightforward and can be read off by using $D_a = \partial_a - i A_a$:
\begin{align}
\delta S[\Phi,X_1,X_2] = \int \td t\td^{d|2}y\, \sqrt{g}\, &\Big[i \Phi^{\dagger}(- i \delta A_0)\Phi -  X_1^\dagger(- i \bar{e}^i \delta A_i \Phi) - X_1(i e^i \delta A_i \Phi^\dagger) \nonumber \\
&+ X_2^\dagger(- i \bar{e}^p \delta A_p\Phi)+ (i e^p \delta A_p\Phi^\dagger) X_2 + \cdots \Big]\,,
\end{align}
where variations with respect to other variables are contained in the dots. This leads to the currents written in~\eqref{susy-j}:
\be
\rho = \Phi^{\dagger} \Phi\,, \quad J^{i} = i \bar{e}^i X_{1}^{\dagger} \Phi - i e^i \Phi^{\dagger} X_1\, \quad J^{p} = -i \bar{e}^p X_{2}^{\dagger} \Phi + i e^p \Phi^{\dagger} X_2 \:.
\ee
For the stress-tensor, we need to vary the action~\eqref{susy-X1X2-1} with respect to the metric, which does not appear explicitly. Hence, we should first reinstate the metric by inserting the identity: $1 = 2e^i \bar{e}_i = 2 \bar{e}^i e_i = 2e^p \bar{e}_p = - 2 \bar{e}^p e_p$. Then we can replace the Zweibeine by the metric using $g^{ij} = 2 (e^i \bar{e}^j + \bar{e}^i e^j), g^{pq} = 2 (\bar{e}^p e^q - e^p \bar{e}^q)$.
\begin{align}
S[\Phi,X_1,X_2]&=\int \td t\td^{d|2}y\, \sqrt{g}\, \Big[ - X_1^\dagger (2 e^j \bar{e}_j)(\bar{e}^iD_i\Phi) - (2 \bar{e}^j e_j) (e^i \widetilde{D}_i\Phi^\dagger)X_1  \nonumber \\
&+ (2 e^q \bar{e}_q) X_2^\dagger(\bar{e}^p D_p\Phi)+ (-2 \bar{e}^q e_q) (e^p \widetilde{D}_p\Phi^\dagger )X_2 + \cdots \Big] \nonumber \\
&= \int \td t\td^{d|2}y\, \sqrt{g}\, \Big[ - g^{ij} X_1^\dagger \bar{e}_j D_i\Phi - g^{ij} e_j (\widetilde{D}_i\Phi^\dagger) X_1 \nonumber \\
&+ g^{pq} \bar{e}_q  X_2^\dagger D_p\Phi + g^{pq} e_q (\widetilde{D}_p\Phi^\dagger) X_2 + \cdots \Big]\:.
\end{align}
We have suppressed other terms in the action. Now we can vary the action with respect to $g^{ab}$, remembering that we should also vary $\sqrt{g} \equiv \sqrt{\text{det }g}$: 
\begin{align}
    T_{ij} &= 2 \frac{\delta S}{\delta g^{ij}} = -2 \bar{e}_j X^{\dagger}_1 D_i \Phi - 2 e_j (\widetilde{D}_i \Phi^{\dagger}) X_1 - g_{ij} \Big[i\Phi^{\dagger}D_0 \Phi -  X_1^\dagger(\bar{e}^iD_i\Phi) -  X_1(e^i \widetilde{D}_i\Phi^\dagger) \nonumber \\
    &+  X_2^\dagger(\bar{e}^p D_p\Phi)+ (e^p \widetilde{D}_p\Phi^\dagger) X_2 + \frac{m}{2} (X_1^\dagger X_1- X_2^\dagger X_2)+V(\Phi) \Big] \:, \\
    T_{pq} &= 2 \frac{\delta S}{\delta g^{pq}} = 2 \bar{e}_q X^{\dagger}_2 D_p \Phi + 2 e_q (\widetilde{D}_p \Phi^{\dagger}) X_2 - g_{pq} \Big[i \Phi^{\dagger}D_0 \Phi -  X_1^\dagger(\bar{e}^iD_i\Phi) -  X_1(\bar{e}^iD_i\Phi)^\dagger \nonumber \\
    &+ X_2^\dagger(\bar{e}^p D_p\Phi)+ (e^p \widetilde{D}_p\Phi^\dagger) X_2 + \frac{m}{2} (X_1^\dagger X_1- X_2^\dagger X_2) +V(\Phi)\Big] \:, \\
    T_{ip} &= 2 \frac{\delta S}{\delta g^{ip}} = \frac{1}{m} (\widetilde{D}_{(i} \Phi^{\dagger} )D_{p)} \Phi  \:,
\end{align}
We have set $g^{ip} = 0$. The variation of $\sqrt{g}$ with respect to $g^{ip}$ is proportional to $g_{ip}$ and hence can be set to zero as well. 
Using the LLL conditions~\eqref{eq:lll-sp} and~\eqref{eq:lll-theta}, and taking the limit $m \to 0$, we can rewrite these expressions as follows: 
\begin{align}
    T_{ij} &= -2 \bar{e}_j X^{\dagger}_1 D_i \Phi - 2 e_j (\widetilde{D}_i \Phi^{\dagger}) X_1 - g_{ij} \Big[2 V(\Phi) - \Phi^{\dagger} \frac{\delta V(\Phi)}{\delta \Phi^{\dagger}} -  \frac{\delta V(\Phi)}{\delta \Phi} \Phi  \Big] \:, \\
    T_{pq} &= 2 \bar{e}_q X^{\dagger}_2 D_p \Phi +2 e_q (\widetilde{D}_p \Phi^{\dagger} )X_2 -  g_{pq} \Big[2 V(\Phi) - \Phi^{\dagger} \frac{\delta V(\Phi)}{\delta \Phi^{\dagger}} -  \frac{\delta V(\Phi)}{\delta \Phi} \Phi \Big] \:, \\
    T_{ip} &= \frac{1}{m} (\widetilde{D}_{(i} \Phi^{\dagger}) D_{p)} \Phi \:.
\end{align}
From these expressions we define ``(anti)holomorphic" stress-tensors and currents (in the LLL):  
\begin{align}
    T &\equiv e^i e^j T_{ij} = - X^{\dagger}_1 e^i D_i \Phi \:, \, \quad \qquad \bar{T} \equiv \bar{e}^i \bar{e}^j T_{ij} =   - \bar{e}^i \widetilde{D}_i \Phi^{\dagger} X_1\:, \\
    T^{\text{tr}} &\equiv  (e^i \bar{e}^j + \bar{e}^i e^j) T_{ij} = -  \Big[2 V(\Phi) - \Phi^{\dagger} \frac{\delta V(\Phi)}{\delta \Phi^{\dagger}} - \frac{\delta V(\Phi)}{\delta \Phi} \Phi  \Big]\:, \\
    T_{\theta} &\equiv e^p e^q T_{pq} = e^p X_{2}^{\dagger} D_p  \Phi \:, \qquad \quad \bar{T}_{\theta} \equiv \bar{e}^p \bar{e}^q T_{pq} = -\bar{e}^p (\widetilde{D} \Phi^\dagger) X_2\:, \\
    T^{\text{tr}}_{\theta} &\equiv  (e^p \bar{e}^q - \bar{e}^p e^q) T_{pq} = 
    -   \Big[2 V(\Phi) - \Phi^{\dagger} \frac{\delta V(\Phi)}{\delta \Phi^{\dagger}} - \frac{\delta V(\Phi)}{\delta \Phi} \Phi  \Big]\:. 
\end{align}

\bibliography{./DisorderLLL.bib}
\bibliographystyle{./JHEP}

\end{document}